\documentclass[a4paper,11pt]{article} 
\pdfoutput=1

\usepackage{jcappub}
\usepackage{amsmath}
\usepackage{amsfonts}
\usepackage{amssymb}
\usepackage{mathtools}
\usepackage{graphics}
\usepackage{citesort}
\usepackage{graphicx}
 \usepackage{url}
 \usepackage{xcolor}
\usepackage{soul}
\usepackage[utf8]{inputenc}
\usepackage[normalem]{ulem}
\usepackage{hyperref}

\newcommand{\eref}[1]{Eq.~(\ref{#1})}
\newcommand{\fref}[1]{Fig.~\ref{#1}}
\newcommand{\sref}[1]{Section~\ref{#1}}
\newcommand{\tref}[1]{Table~\ref{#1}}

\newcommand{\Tr}[1]{{\rm Tr}(#1)}
\newcommand{\Pz}[1]{P_{z, {\rm int}, #1}}

\title{
 Neutrino flavor mixing breaks isotropy in the early universe
}

\author[a]{Rasmus S.~L.~Hansen,}
\author[a]{Shashank Shalgar,}
\author[a]{and Irene Tamborra}
\affiliation[a]{Niels Bohr International Academy and DARK, Niels Bohr Institute, University of Copenhagen, Blegdamsvej 17, 2100, Copenhagen, Denmark}
\emailAdd{rslhansen@nbi.ku.dk}
\emailAdd{shashank.shalgar@nbi.ku.dk}
\emailAdd{tamborra@nbi.ku.dk}

\abstract{The neutrino field is commonly assumed to be isotropic and homogeneous in the early universe. 
However, due to the large neutrino density, a small perturbation of the isotropy of the neutrino field could potentially be amplified by the non-linear flavor mixing caused by neutrino self-interactions. We carry out the first numerical simulations of the neutrino flavor evolution in a multi-angle anisotropic setting. Due to the computational challenges involved, we adopt a simplified framework consisting of a homogeneous universe with two angle bins---left and right moving modes---for neutrinos and antineutrinos, together with an approximate form for the collision term which goes beyond the commonly adopted damping approximation. 
 By assuming a small initial left-right asymmetry of $\mathcal{O}(10^{-15})$,
 we convincingly demonstrate that flavor evolution can be affected in both mass orderings, with implications on the effective number of thermally excited neutrino species ($N_{\mathrm{eff}}$). Notably, the correction to $N_{\rm eff}$ is comparable to higher order corrections from finite temperature QED effects in normal ordering.
 In addition, by assuming an initial lepton asymmetry in the neutrino sector of the same order as the baryon one [$\mathcal{O}(10^{-9})$], we find that the neutrino-antineutrino asymmetry grows by several orders of magnitude for isotropic as well as anisotropic initial conditions.
 This work clearly shows that it is imperative to critically revisit standard assumptions concerning neutrino flavor mixing in the early universe, especially in the light of possible implications on the cosmological observables. 
}

\begin{document}
\maketitle

\section{Introduction}

The early universe is rich in neutrinos, hence the coherent forward scattering of neutrinos among themselves (neutrino self-interactions) cannot be neglected. This phenomenon can modify the flavor evolution, similar to the Mikheev-Smirnov-Wolfenstein (MSW) effect~\cite{Mikheev:1986if,1985YaFiz..42.1441M,1978PhRvD..17.2369W} experienced by neutrinos forward scattering off electrons. 
However, unlike the MSW effect, neutrino self-interactions are non-linear in nature; this leads to a very interesting and rich phenomenology in neutrino-dense astrophysical objects as well as in the early universe~\cite{Mirizzi:2015eza,Duan:2010bg,Chakraborty:2016yeg,Tamborra:2020cul}. 
Neutrino self-interactions lead to a ``collective'' evolution of neutrino flavor, i.e.~neutrinos with different momenta evolve in-phase. This phenomenon was first considered by Kostelecky, Pantaleone, and Samuel in the context of the early universe~\cite{Kostelecky:1993yt,Kostelecky:1993ys,Kostelecky:1993dm,Kostelecky:1994dt}.

A large body of work explores the neutrino flavor evolution in the presence of neutrino self-interactions, assuming a large lepton asymmetry in the early universe (see e.g.~\cite{Dolgov:2002ab, Wong:2002fa, Abazajian:2002qx}), i.e.,~in scenarios invoking significant contributions from physics beyond the Standard Model. On the other hand, work focusing on Standard Model physics disregards neutrino self-interactions~\cite{Mangano:2005cc, deSalas:2016ztq,Froustey:2020mcq, Bennett:2020zkv}, relying on the argument that neutrinos and antineutrinos have almost equal abundances and therefore the neutrino-neutrino term cancels out in the Hamiltonian. 
We stress that, due to the non-linear nature of the neutrino flavor evolution, flavor mixing in the early universe cannot be neglected a priori, see e.g.~Refs.~\cite{Hasegawa:2019jsa,Johns:2016enc,Sawyer:2020goq}. 

The non-linear nature of neutrino flavor evolution can lead to spontaneous breaking of the symmetries that may exist in a system~\cite{Raffelt:2013rqa,Mangano:2014zda, Duan:2014gfa}.
This phenomenon has been widely studied in the context of the neutrino flavor evolution in compact astrophysical objects, but has been ignored in the early universe. 
 However, in the linear stability analysis performed in Ref.~\cite{Cirigliano:2017hmk}, it was found that homogeneity and isotropy are spontaneously broken in the presence of a large lepton asymmetry.
It was also concluded that collisions in the early universe are unlikely to suppress the occurrence of symmetry breaking effects. 

A self-consistent estimation of the neutrino flavor evolution in the early universe in the presence of collisions is extremely challenging, even under the simplifying assumptions of homogeneity and isotropy, and has not been performed to date.
For this reason, most of the existing literature focuses on solving the Boltzmann equations for the neutrino gas, while considering a damping approximation for the non-forward collisions~\footnote{Terms beyond damping are included in Ref.~\cite{deSalas:2016ztq,Froustey:2020mcq,Bennett:2020zkv}, but the neutrino self-interaction potential is neglected.}.

In order to demonstrate the limitations induced in our understanding of the flavor conversion physics by the isotropy assumption, we solve the full set of momentum dependent non-linear equations of motion of neutrinos and antineutrinos by allowing for anisotropic solutions. In order to tackle this problem numerically, we rely on a simplified model of a homogeneous universe with two angle bins for neutrinos and antineutrinos, left ($L$) and right ($R$) moving (anti)neutrinos. The isotropic case is recovered when the flavor evolution along $L$ is exactly equivalent to the one along $R$.
In addition, we only consider neutrino flavor conversions between electron neutrinos ($\nu_e$) and a linear combination of muon and tau neutrinos ($\nu_x$). The third neutrino state, $\nu_y$, is only included when calculating the expansion rate of the universe. We also adopt an approximated collision term where anisotropies can be accounted for. In the homogeneous and isotropic case, the full collision integral can be reduced analytically to a two-dimensional integral, which can be evaluated numerically. However, this is still numerically expensive, and the analytical reduction is not possible when the assumption of isotropy is relaxed.
We demonstrate very convincingly that the assumptions of homogeneity and isotropy for the neutrino field are not justified and that a sustained effort is required by the community to finally gauge the role of neutrino flavor conversions in the early universe.

This paper is structured as follows. The equations of motion are introduced in Sec.~\ref{sec:eom} together with the collision term. In Sec.~\ref{sec:linear}, we consider a  simplified set of equations of motion and perform a momentum dependent linear stability analysis of our homogeneous model with two angle bins. We determine the growth rates as well as the symmetries of the unstable modes in the absence of collisions. Our non-linear numerical results are presented in Secs.~\ref{sec:isotropic} and \ref{sec:anisotropic}, where the assumption of isotropy is relaxed and collisions are taken into account. Section~\ref{sec:symmbreaking} explores the flavor conversion physics for anisotropic initial conditions and in the presence of a neutrino-antineutrino asymmetry. A discussion on the implications of our findings on cosmological observables, such as the effective number of radiation species is reported in Sec.~\ref{sec:discussion}. Finally, our findings are summarized in Sec.~\ref{sec:conclusions}. The collision term adopted in this work is reported in Appendix~\ref{app:collterm}, while details on the numerical implementation of the equations of motion can be found in Appendix~\ref{app:QKE}.
In order to corroborate the robustness of our numerical findings, Appendix~\ref{app:nunubar} explores the  growth of the neutrino-antineutrino asymmetry in a very simple model and by relying on the linear stability analysis.

\section{Tracking flavor conversions in the early universe}
\label{sec:eom}

 In the early universe, neutrinos are in thermal equilibrium for temperatures above $\mathcal{O}(10)$~MeV.
However, as the temperature decreases due to the expansion of the universe, weak decoupling begins at $\simeq 1$~MeV, and the neutrino energy distributions deviate from their thermal shape; it is then possible for flavor conversions to be triggered. Favorable conditions for neutrino flavor evolution persist down to $\mathcal{O}(0.3)$~MeV where the Big Bang nucleosynthesis starts. During this time, spanning two orders of magnitude in temperature, the neutrino vacuum frequency is inversely proportional to the temperature, while the collision strength varies as the fifth power of the temperature. This results in a large variation in the time-scale of the flavor evolution with respect to the one that governs collisions. In this section, after introducing the equations of motion, we outline the main ingredients of our collision term and the framework of our homogeneous universe with two angle bins.

\subsection{Equations of motion}
As the temperature decreases and the neutrino distribution deviates from the thermal equilibrium one, electrons and positrons become non-relativistic at temperatures just below $\sim 1$~MeV.
 We assume that photons, electrons and positrons are initially in equilibrium at all times with a temperature $T_\gamma$.
As electrons and positrons become non-relativistic, $T_\gamma$ will deviate from the comoving temperature $T_{\rm cm}$ which is defined to be inversely proportional to the scale factor. Since neutrinos are in the process of decoupling, they will also be heated by a small transfer of entropy. This entropy transfer can be followed by solving the quantum kinetic equations (QKEs). 

To follow the evolution of $T_\gamma$, we rely on the continuity equation
\begin{equation}
 \label{eq:EconsT}
 \frac{dT_\gamma}{dt} = \frac{-4 H(\varrho_\gamma + \varrho_\nu) - 3 H(\varrho_e + P_e) - \frac{d\varrho_\nu}{dt}}{\frac{\partial\varrho_\gamma}{\partial T_\gamma} + \frac{\partial \varrho_e}{\partial T_\gamma}}\ ,
\end{equation}
\begin{figure}[tbp]
 \centering
 \includegraphics[width=0.7\textwidth]{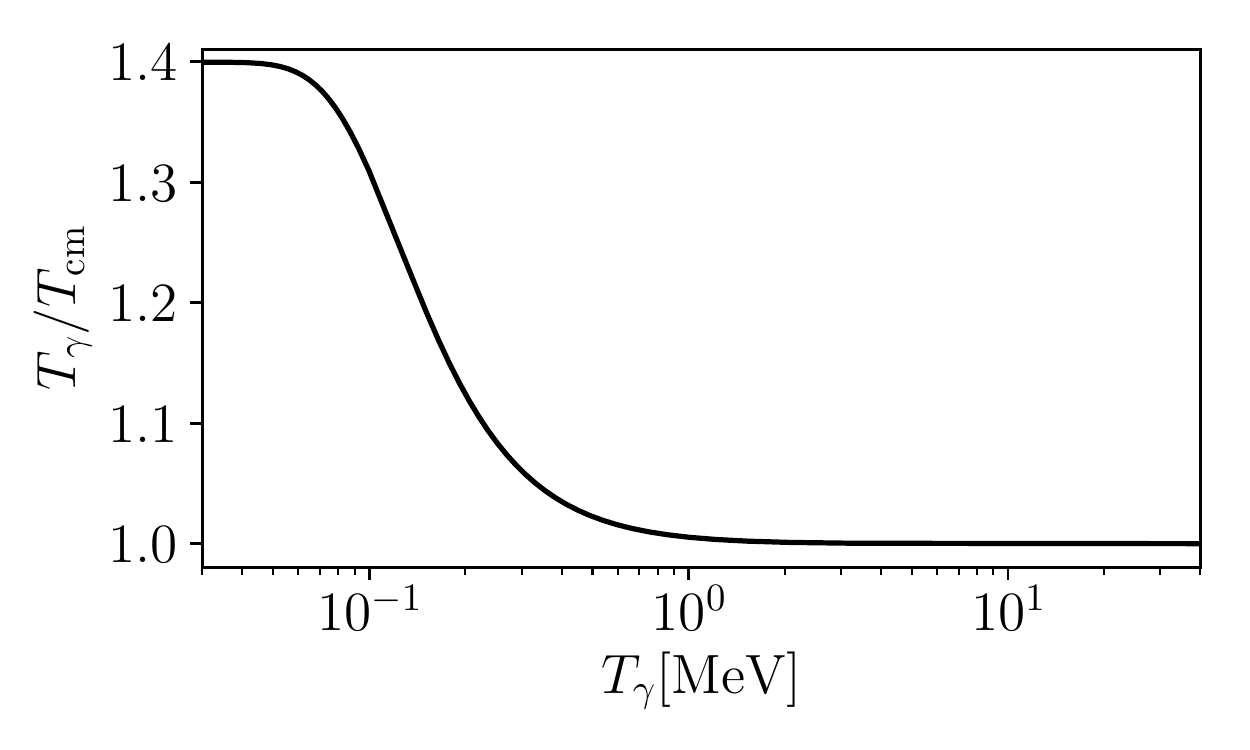}\\
 \caption{Ratio of the photon temperature to the comoving temperature as a function of the photon temperature. The photons are heated below $1$~MeV by annihilation of electrons and positrons.}
 \label{fig:TcmTgamma}
\end{figure}%
where $\varrho_{x}$ and $P_x$, with $x= \gamma, e, \nu$, are the density and the pressure of photons, electrons, and neutrinos, respectively. 
The Hubble parameter $H$ is given by the Friedmann equation
\begin{equation}
 \label{eq:Friedmann}
 H \equiv \frac{\dot{a}}{a} = \sqrt{\frac{8\pi G \varrho}{3}} = \sqrt{\frac{8\pi\varrho}{3}} \frac{1}{m_{\rm Pl}}\ ,
\end{equation}
where $a$ is the scale factor, $\varrho$ is the total density, and $m_{\rm Pl}$ is the Planck mass.
Density, pressure, and their derivatives for photons and $\nu_y$ are calculated analytically. For electrons and positrons, instead, $\varrho_{x}$ and $P_{x}$ are calculated numerically by taking the electron mass into account; for $\nu_e$ and $\nu_x$, results from the QKEs are used.
The equations are integrated as functions of the scale factor $a$, by relying on $\dot{a} = da/dt = H a$. The evolution of $T_\gamma/T_{\rm cm}$ is shown in \fref{fig:TcmTgamma}, where the photon heating due to electron-positron annihilation is visible between $1$~MeV and $0.03$~MeV. 

The neutrino flavor evolution is tracked by means of the density matrix formalism~\cite{Dolgov:1980cq, Stodolsky:1986dx, Barbieri:1989ti, Sigl:1992fn}. Since we intend to allow for inhomogeneous and anisotropic solutions, the density matrix $\rho$ depends on the three-dimensional momentum $\mathbf{p}$ with magnitude $p$ as well as the position $\mathbf{x}$. The diagonal entries in $\rho(\mathbf{p},\mathbf{x})$ give the densities of each neutrino flavor while the off-diagonal terms encode the coherence information among neutrinos~\footnote{We write the density matrices in terms of creation and annihilation operators as $\rho_{ij} = \left\langle a_j^\dagger a_i\right\rangle$ and $\bar{\rho}_{ij} = \left\langle b_j^\dagger b_i\right\rangle$, whereas the opposite order of the indices in $\bar{\rho}$ is often used when studying supernova neutrinos.}. The flavor evolution is described by the following equation~\cite{Sigl:1992fn, Blaschke:2016xxt}:
\begin{equation}
 \label{eq:rhoeom}
 \frac{d \rho(\mathbf{p},\mathbf{x})}{dt} = -i \left[ \mathcal{H}(\rho,\mathbf{p},\mathbf{x}), \rho(\mathbf{p},\mathbf{x})\right] + \mathcal{C}(\rho,\mathbf{p},\mathbf{x})\ ,
\end{equation}
 where $\mathcal{H}$ is the Hamiltonian that governs the flavor conversion physics due to the vacuum term as well as the forward scattering of neutrinos due to the fermionic background,
while $\mathcal{C}$ is the collision term accounting for momentum-changing interactions. In our framework of two neutrino flavors, the density matrices are
\begin{equation}
 \label{eq:rho}
 \rho =
 \begin{pmatrix}
 \rho_{ee} & \rho_{ex} \\ \rho_{x e} & \rho_{xx}
 \end{pmatrix}\ \qquad \mathrm{and} \qquad \ 
 \bar{\rho} =
 \begin{pmatrix}
 \bar{\rho}_{ee} & \bar{\rho}_{ex} \\ \bar{\rho}_{x e} & \bar{\rho}_{xx}
 \end{pmatrix}\ .
\end{equation}
Note that, although the two flavor framework does not capture all the features present in the full three flavor system, we expect that the introduction of anisotropic solutions will have a similar impact in both cases.

The Hamiltonian is given by~\cite{Notzold:1987ik, Sigl:1992fn}
\begin{align}
 \mathcal{H}(\rho,\mathbf{p},\mathbf{x}) &= \frac{\mathcal{U} \mathcal{M}^2 \mathcal{U}^\dagger}{2 p} + \sqrt{2} G_{
\rm F} \int \frac{d^3\mathbf{p}'}{(2\pi)^3} (\rho(\mathbf{p}',\mathbf{x}) - \bar{\rho}^*(\mathbf{p}',\mathbf{x}))(1 - \mathbf{v}' \cdot \mathbf{v})\nonumber\\
 &\quad- \frac{8\sqrt{2} G_{
\rm F} p}{4} \left( \frac{\mathcal{E}_l+\frac{1}{3}\mathcal{P}_l}{m_W^2} + \frac{1}{m_Z^2} \int \frac{d^3\mathbf{p}'}{(2\pi)^3} p' \; (\rho(\mathbf{p'},\mathbf{x}) + \bar{\rho}^*(\mathbf{p'},\mathbf{x}))(1 - \mathbf{v}' \cdot \mathbf{v})^2\right)\ . \label{eq:Hamiltonian}
\end{align}
Here the first term describes vacuum mixing, the second term is proportional to the neutrino asymmetry, while the third term accounts for non-local gauge boson effects. The non-local gauge boson effects due to neutrinos are negligible in our case since the abundances of $\nu_e$ and $\nu_x$ are similar in contrast to the abundances of the charged leptons. Still, we include the effect in the numerical calculations for completeness. The Hamiltonian for antineutrinos is obtained by substituting $\rho \rightarrow \bar{\rho}$.

In the first term of \eref{eq:Hamiltonian}, $\mathcal{M}^2$ is a diagonal matrix with the squares of the neutrinos masses as its entries, while $\mathcal{U}$ is the Pontecorvo-Maki-Nakagawa-Sakata matrix.
In the second term, $G_{\rm F}$ is the Fermi constant, and $\mathbf{v}=\mathbf{p}/p$ is the velocity vector (we assume ultra-relativistic neutrinos, $|\mathbf{v}| = c = 1$).
In the last term, $m_W$ and $m_Z$ denote the masses of the $W$ and $Z$ bosons, respectively. 
$\mathcal{E}_l$ and $\mathcal{P}_l$ are diagonal matrices with the total energy density and pressure densities, respectively, with a contribution from charged leptons and antileptons. 
In the presence of bulk flows in the background, additional terms are present in the Hamiltonian~\cite{Vlasenko:2013fja,Cirigliano:2014aoa,Vlasenko:2014bva}. However, we neglect such terms in this work due to the approximate isotropy of the background.

We quantify the effect of oscillations by tracking the total neutrino number density for each flavor
\begin{equation}
 \label{eq:nalpha}
 n_{\alpha} = \int\frac{dp}{2\pi^2} p^2 \rho_{\alpha\alpha}\ .
\end{equation}
In addition, we are interested in the absolute change in number distributions due to oscillations and heating from electron-positron annihilation
\begin{equation}
 \label{eq:Deltanabs}
 \Delta \left(\frac{dn}{dr}\right) = \frac{T_{\rm cm}}{2\pi^2} p^2 (\rho_{\alpha\alpha} - f_0)\ .
\end{equation}
In both cases, we normalize these quantities by
\begin{equation}
 \label{eq:n0}
 n_0 = \int\frac{dp}{2\pi^2} p^2 f_0(p)\ 
\end{equation}
such that the dependence on the comoving temperature drops out.

\subsection{The collision term}

The collision term in \eref{eq:rhoeom} accounts for non-forward scattering interactions of neutrinos with other neutrinos in the medium as well as charged leptons and nucleons in the plasma. Given the tiny value of the baryon asymmetry, the density of nucleons is extremely small at the relevant temperatures and, hence, we will neglect contributions from nucleons. 
The full collision term requires the evaluation of nine-dimensional momentum integrals at each point in space. In the isotropic and homogeneous case, seven of the integrals can be solved analytically~\cite{Dolgov:1997mb, Hannestad:1995rs, HahnWoernle:2009qn}, and only two must be performed numerically. However, this is not possible in the general anisotropic case. Furthermore, even just evaluating the two integrals is numerically very expensive~\cite{Gariazzo:2019gyi,Akita:2020szl,Grohs:2020xxd}, hence we choose to use an approximated collision term. 

 In the collision term, we separately consider the scattering with electrons and positrons, annihilations to electrons and positrons, as well as neutrino-neutrino and neutrino-antineutrino scatterings, as summarized in \tref{tab:collisions}. A derivation of our collision term is reported in Appendix~\ref{app:collterm}, where we also verify that our approximation well reproduces the results from the full collision term.

\begin{table}[tbp]
 \centering
 \caption{Reactions included in the collision term and the corresponding rates in Eqs.~(\ref{eq:Cee}) and (\ref{eq:Cmm}). The neutrino indices are $\alpha,\beta \in \left\{e,x\right\}$.}
 \begin{tabular}{l c l }
 \hline\hline
 Process & Reaction & Rate\\
 \hline
 Scattering & $\nu_\alpha e^\pm \leftrightarrow \nu_\alpha e^\pm$ & $\Gamma_{s,\alpha}$\\
 Annihilation & $\nu_\alpha \bar{\nu}_\alpha \leftrightarrow e^- e^+$ & $\Gamma_{a,\alpha}$ \\
 Neutrino-neutrino scattering & $\nu_\alpha \nu_\beta \leftrightarrow \nu_\alpha \nu_\beta$ & $\Gamma_{\nu\nu}$\\[2mm]
 Neutrino-antineutrino collisions & $\begin{matrix} \nu_\alpha \bar{\nu}_\beta \leftrightarrow \nu_\alpha \bar{\nu}_\beta, \; \alpha \neq \beta\\ \nu_\alpha \bar{\nu}_\alpha \leftrightarrow \nu_\beta \bar{\nu}_\beta\end{matrix}$ & $\Gamma_{\nu\bar{\nu}}$ \\\hline\hline
 \end{tabular}
 \label{tab:collisions}
\end{table}

We assume that all distribution functions have no chemical potential when calculating the collision rates. The functional form of the approximate collision term is determined by including one equilibrium distribution function in the gain term, while all other neutrino distribution functions are represented by normalized energy densities, $u_{\alpha\beta}$. In order to accurately reproduce the behavior of the different terms, we appropriately define the equilibrium function for each process in terms of temperature and (pseudo-)chemical potential and neglect Pauli blocking. In the homogeneous and isotropic case, this leads to the following diagonal terms:
\begin{align}
 \mathcal{C}_{ee} &= \Gamma_{s,e} \left(f(T_\gamma,\pi_{e}) - \rho_{ee}\right) + \Gamma_{a,e} \left( \left(\tfrac{T_\gamma}{T_{\rm cm}}\right)^4 f(T_\gamma,\mu_{e}) - \rho_{ee} \right) - \Gamma_G {\rm Re}\left( \bar{u}_{ex} \rho_{ex}^* \right) \nonumber\\
 &\quad+ (\Gamma_{\nu\nu} (2u_{ee} + u_{xx}) + \Gamma_{\nu\bar{\nu}}( 4 \bar{u}_{ee} + \bar{u}_{xx}) ) (f_{\nu_e} - \rho_{ee}) \nonumber\\
 &\quad + \Gamma_{\nu\bar{\nu}} (\bar{u}_{xx} f_{\nu_x} - \bar{u}_{ee}\rho_{ee}) + {\rm Re}((\Gamma_{\nu\nu} u_{ex}^* + 4 \Gamma_{\nu\bar{\nu}} \bar{u}_{ex}^*)(f_{ex} - \rho_{ex}))\ ,\label{eq:Cee}
\end{align}
\begin{align}
 \mathcal{C}_{xx} &= \Gamma_{s,x} (f(T_\gamma,\pi_{x}) - \rho_{xx}) + \Gamma_{a,x} \left( \left(\tfrac{T_\gamma}{T_{\rm cm}}\right)^4 f(T_\gamma,\mu_{x}) - \rho_{xx}\right)- \Gamma_G {\rm Re}\left( \bar{u}_{ex} \rho_{ex}^* \right)\nonumber\\
 &\quad+(\Gamma_{\nu\nu} (u_{ee} + 2 u_{xx}) + \Gamma_{\nu\bar{\nu}} (4 \bar{u}_{xx} + \bar{u}_{ee}) )(f_{\nu_x} -\rho_{xx})\nonumber\\ 
 &\quad + \Gamma_{\nu\bar{\nu}} (\bar{u}_{ee} f_{\nu_e} - \bar{u}_{xx} \rho_{xx}) + {\rm Re}((\Gamma_{\nu\nu}u_{ex}^* + 4\Gamma_{\nu\bar{\nu}}\bar{u}_{ex}^*)(f_{ex} - \rho_{ex}))\ ;\label{eq:Cmm}
\end{align}
the scattering of neutrinos with electrons and positrons gives rise to the terms proportional to $\Gamma_{s,\alpha}$; these terms depend on the deviation from equilibrium, and conserve the number density of each particle. Annihilations of neutrinos and antineutrinos to electrons and positrons lead to terms depending on $\Gamma_{a, \alpha}$ that are functions of the deviation from equilibrium and conserve the lepton number for each neutrino flavor. The factor $(T_\gamma/T_{\rm cm})^4$ accounts for the temperature difference between electrons and neutrinos. The additional diagonal contribution proportional to $\Gamma_G$ accounts for the annihilation of neutrinos and antineutrinos in a flavor coherent state. Apart from the term containing $\Gamma_G$, the form of the collision term resembles the one of a Boltzmann equation in the relaxation time approximation~\cite{Johns:2019hjl}. This is not surprising since our approximation is designed to ensure the principle of detailed balance is not violated; the same as in the relaxation time approximation. 
Neutrino-neutrino (neutrino-antineutrino) scatterings give rise to terms with $\Gamma_{\nu\nu}$ ($\Gamma_{\nu\bar{\nu}}$). The second lines of \eref{eq:Cee} and \eref{eq:Cmm} are simple relaxation terms that depend on the deviation from equilibrium. The first term of the third line in both equations is due to one neutrino flavor annihilating into the other flavor and pushes towards flavor equilibrium. The very last term in both equations is the analog of the $\Gamma_G$ term, where $f_{ex}$ accounts for the gain from other momentum states while $\rho_{ex}$ is the corresponding loss term.

The off-diagonal collision term is
\begin{align}
 \mathcal{C}_{ex} &= - D \rho_{ex} + d \;f_{ex} - C \bar{u}_{ex} (\rho_{ee} + \rho_{xx})\nonumber\\
 &\quad+ \Gamma_{\nu\bar{\nu}} (\bar{u}_{ee} + \bar{u}_{xx}) (2f_{ex} - 3\rho_{ex})+ \Gamma_{\nu\bar{\nu}}\bar{u}_{ex} ((f_{\nu_e} + f_{\nu_x}) - 2(\rho_{ee} + \rho_{xx}))\nonumber\\
 &\quad + \tfrac{3}{2} \Gamma_{\nu\nu} (u_{ee} + u_{xx}) ( f_{ex} - \rho_{ex}) + \tfrac{1}{2}\Gamma_{\nu\nu} u_{ex} ((f_{\nu_e}+f_{\nu_x}) - (\rho_{ee} + \rho_{xx}))\ .\label{eq:Cem}
\end{align}
Annihilations also contribute to the damping term $D$ and, in addition, there is a term proportional to $C$ that depends on the coherence of antineutrinos which cannot be explained heuristically~\cite{McKellar:1992ja}.

The equilibrium distributions are given by Fermi-Dirac distributions:
 \begin{equation}
 \label{eq:fTmu}
 f(T_{\rm eq},\mu) = \frac{1}{\exp(p/T_{\rm eq}-\mu/T_{\rm eq}) + 1}\ ,
\end{equation}
$T_{\rm eq}$ is a representative temperature which can be $T_{\rm cm}$, $T_\gamma$ or the neutrino temperature $T_{\nu_\alpha}$ which will be introduced below. The other parameter $\mu$ is the chemical potential ($\mu_{\bar{x}} = - \mu_x$) or the pseudo-chemical potential ($\pi_x$). The chemical potentials are such that lepton number is conserved. The pseudo-chemical potentials for electron and positron interactions are defined so that that the number density of each neutrino flavor (and hence also the lepton number) is conserved, while the energy density of each neutrino flavor is not conserved. For neutrino-neutrino interactions, the pseudo-chemical potential and the temperature $T_{\nu_\alpha}$ are determined to conserve both number and energy density of each neutrino flavor simultaneously~\footnote{Terms of the kind $(\bar{u}_{\alpha\alpha} f_{\nu_\alpha} - \bar{u}_{\beta\beta}\rho_{\beta\beta})$ do not conserve number or energy. However, the sum $\mathcal{C}_{\alpha\alpha}+\mathcal{C}_{\beta\beta}$ conserves both the total number- and energy densities. Similarly, $\mathcal{C}_{\alpha\alpha}-\bar{\mathcal{C}}_{\alpha\alpha}$ conserves the lepton number.}. We use $f_0 = f(T_{\rm cm},0)$. For neutrinos, $f_{\nu_\alpha} = f(T_{\nu_\alpha}, \pi_{\nu_\alpha})$. The off-diagonal part of the density matrices is such that 
\begin{equation}
 \label{eq:femu}
 f_{ex} = (a_x +b_x p) f_0 + i (a_y + b_y p) f_0\ , 
\end{equation}
where $a_x$, $b_x$, $a_y$ and $b_y$ are determined so to conserve the first and second moments of $\rho_{ex}$.
The normalized energy densities are
\begin{equation}
 \label{eq:ualphabeta}
 u_{\alpha\beta} = \frac{\int dp \; p \; \rho_{\alpha\beta}}{\int dp \; p f_0}\ ;
\end{equation}
 their appearance ensures detailed balance for neutrino collisions. 
The front factors $\Gamma_{s,\alpha}$, $\Gamma_{a,\alpha}$, $\Gamma_G$, $\Gamma_{\nu\nu}$, $\Gamma_{\nu\bar{\nu}}$, $D$, $d$, and $C$ all have the form
\begin{equation}
\Gamma_x (D,d,C) = y_{x} G_{
\rm F}^2 p T_{\rm cm}^4\label{eq:Gammax}\ \qquad \mathrm{with}\ \qquad x \in \{ s,\alpha ; a,\alpha ; \nu\nu ; \nu\bar{\nu}; G ; D ; d ; C \}\ ,
\end{equation}
where the coefficients are calculated in Appendix~\ref{app:collterm}.

\subsection{Homogeneous universe model with two angle bins}
The collision term introduced in Eqs.~(\ref{eq:Cee})-(\ref{eq:Cem}) assumes homogeneity and isotropy. 
Since inhomogeneity implies anisotropy, but the converse is not necessarily true, the simplest self-consistent choice is to take a homogeneous universe with well defined $L$ and $R$ directions. To this purpose, we introduce left and right moving modes denoted by the subscripts $L$ and $R$, respectively. 
Then, Eq.~(\ref{eq:rhoeom}) becomes
\begin{align}
 \label{eq:qke1d}
 \frac{\partial\rho_R(p)}{\partial t} - H p \frac{\partial \rho_R(p)}{\partial p} = -i[\mathcal{H}_R(\rho_R, \rho_L, p),\rho_R(p)] + \mathcal{C}_R(\rho_R, \rho_L,p)\ ,\\
 \frac{\partial\rho_L(p)}{\partial t} - H p \frac{\partial \rho_L(p)}{\partial p} = -i[\mathcal{H}_L(\rho_R, \rho_L, p),\rho_L(p)] + \mathcal{C}_L(\rho_R, \rho_L,p)\ ,
 \label{eq:qke1d2}
\end{align}
where $p$ is now a scalar corresponding to the neutrino energy, with direction defined by $L$ and $R$.
The Hamiltonian for the right moving mode is
\begin{align}
 \mathcal{H}_R(\rho_R,\rho_L,p) &= \frac{\mathcal{U} \mathcal{M}^2 \mathcal{U}^\dagger}{2 p} + \sqrt{2} G_{
\rm F} \int \frac{dp'}{2\pi^2} (\rho_L(p') - \bar{\rho}^*_L(p')) - \frac{8\sqrt{2} G_{
\rm F} p}{3} \frac{\mathcal{E}_l}{m_W^2}\ , \label{eq:1DHamiltonian}
\end{align}
where we notice that there is no contribution in the self-interaction term of $\mathcal{H}_R$ from $\rho_R$ and $\bar{\rho}^*_R$. The Hamiltonian for the left moving mode is obtained by $R \leftrightarrow L$. 
The collision term is outlined in Appendix~\ref{app:collterm} where, while calculating the collision rates, we maintain the assumption of isotropy, but we introduce anisotropy for the equilibrium distributions.

Having verified that the diagonal part of the collision term is working well (see Appendix~\ref{app:collterm}) and after introducing our anisotropic model, the next step is to explore the flavor conversion physics. However, before we proceed to solve the full set of equations, it is instructive to consider a set of simplified equations and perform a linear stability analysis, which will be crucial in interpreting the numerical results.

\section{Linear stability analysis}
\label{sec:linear}

The linear stability analysis~\cite{Banerjee:2011fj} can help us in understanding the conditions under which flavor conversions set in. In this section, we employ the linear stability analysis, but neglect the collision term for simplicity. In fact, Ref.~\cite{Cirigliano:2017hmk} found that collisions should have little impact on the linear regime for the temperature range of interest.
Similarly, we neglect the effect of Hubble expansion and the non-local gauge boson effect from neutrinos.
Neglecting the Hubble expansion also implies that there is no difference between $T_{\rm cm}$ and $T_\gamma$. Hence we use $T_{\rm cm}$ throughout this section. 
In order to avoid constant terms in the final linear equations, we use a vanishing mixing angle for the linear analysis. 

With these assumptions,  Eqs.~(\ref{eq:qke1d}) and (\ref{eq:qke1d2}) become
\begin{align}
 \frac{\partial\rho_R(p)}{\partial t} &= - i [\mathcal{H}_{\rm vac}(p) + \mathcal{H}_{\rm m}(p) + \mathcal{H}_{\nu\nu,R},\rho_R(p)],\label{eq:qkesimple1}\\
 \frac{\partial\bar{\rho}_R(p)}{\partial t} &= - i [\mathcal{H}_{\rm vac}(p) + \mathcal{H}_{\rm m}(p) - \mathcal{H}_{\nu\nu,R}^*,\bar{\rho}_R(p)]\ ,\\
 \frac{\partial\rho_L(p)}{\partial t} &= - i [\mathcal{H}_{\rm vac}(p) + \mathcal{H}_{\rm m}(p) + \mathcal{H}_{\nu\nu,L},\rho_L(p)],\\
 \frac{\partial\bar{\rho}_L(p)}{\partial t} &= - i [\mathcal{H}_{\rm vac}(p) + \mathcal{H}_{\rm m}(p) - \mathcal{H}_{\nu\nu,L}^*,\bar{\rho}_L(p)]\ ,\label{eq:qkesimple4}
\end{align}
where
\begin{align}
 \mathcal{H}_{\rm vac} &= \omega(p) \; \mathrm{diag}(-\tfrac{1}{2},\tfrac{1}{2}) = \frac{\Delta m^2}{2 p} \mathrm{diag}(-\tfrac{1}{2},\tfrac{1}{2})\ ,\label{eq:Hvac}\\
 \mathcal{H}_{\rm m} &= \lambda(p) \; \mathrm{diag}(-\tfrac{1}{2},\tfrac{1}{2}) = \frac{14\sqrt{2}G_{
\rm F} p \pi^2 T_{\rm cm}^4}{45 m_W^2} \mathrm{diag}(-\tfrac{1}{2},\tfrac{1}{2})\ ,\label{eq:Hm}\\
 \mathcal{H}_{\nu\nu,R} &= \mu \int dr' r'^2 (\rho_L(r') - \bar{\rho}_L(r')^*) = \frac{G_{\rm F} T_{\rm cm}^3}{\sqrt{2}\pi^2} \int dr' r'^2 (\rho_L(r') - \bar{\rho}_L(r')^*) \; . \label{eq:Hnu}
\end{align}
and we express the integrals in terms of comoving momentum $r' = p'/T_{\rm cm}$.
The expressions for left moving modes are obtained by $R \leftrightarrow L$. Note that the $\nu$-$\nu$ potential, $\mu$, has received a factor of 2 from $(1-\mathbf{v}\cdot\mathbf{v}')$ and a factor of $\frac{1}{2}$ from the angular integral which only runs over  half hemisphere. To simplify the equations, we have defined $\mathcal{H}_{\rm vac}$ and $\mathcal{H}_{\rm m}$ such that they are traceless. 

The simplified equations in Eqs.~(\ref{eq:qkesimple1})--(\ref{eq:qkesimple4}) bear some similarity to other models in the literature~\cite{Raffelt:2013rqa,Duan:2013kba,Hansen:2014paa,Chakraborty:2016lct,Capozzi:2016oyk,Martin:2019dof,Mirizzi:2015fva}; in particular to the model considered in Ref.~\cite{Raffelt:2013rqa} for which the authors demonstrate that self-induced flavor conversions can take place for both neutrino mass orderings.
In \cite{Raffelt:2013rqa}, the matter term is neglected and a single energy approximation is used. With these simplifications, it is found that symmetric initial conditions lead to stability for what concerns flavor conversions for normal mass ordering (NO) where $\omega>0$. For inverted mass ordering (IO) where $\omega<0$, the system is unstable and exhibits the well-known bipolar oscillations.
If the initial conditions are instead antisymmetric in left and right moving modes, the cases of NO and IO are exchanged; that is, NO is unstable while IO is stable.

Although the above considerations are for a much simpler system than the one set up to describe the early universe, we still expect some of the features to be present. First, we expect little flavor conversion in NO when left-right symmetry is assumed, while flavor conversions should take place for IO even in a symmetric setup. Second, we expect flavor conversions in NO to be triggered when we break left-right symmetry.
This is also consistent with the predictions made in Ref.~\cite{Cirigliano:2017hmk}.

In the following, we make the predictions more quantitative by including the matter term and performing a linear stability analysis.
In order to linearize the equations, we write the density matrices as
\begin{align}
  \rho_R(p) &= \frac{1}{2} \Tr{\rho_R(p)} +  \frac{1}{2}
  \begin{pmatrix} s_p & \epsilon_{R,p} \\ \epsilon^*_{R,p} & -s_p\end{pmatrix}, \\
  \bar{\rho}_R(p) &= \frac{1}{2} \Tr{\bar{\rho}_R(p)} + \frac{1}{2}
  \begin{pmatrix} \bar{s}_p & \bar{\epsilon}^*_{R,p} \\ \bar{\epsilon}_{R,p} & -\bar{s}_p\end{pmatrix}\ ,
\end{align}
and similarly for $\rho_L(p)$ and $\bar{\rho}_L(p)$. We assume that the initial state is very close to being isotropic, and consequently $s_p$ and $\bar{s}_p$ do not depend on the angular variable. Assuming $\epsilon_{R,p}$, $\epsilon_{L,p}$, $\bar{\epsilon}_{R,p}$, and $\bar{\epsilon}_{L,p} \ll s_p, \bar{s}_p$,  we can ignore terms of $\mathcal{O}(\epsilon^{2})$ and higher, resulting in the following equations:
\begin{align}
  \label{eq:epsilon}
  \frac{\partial \epsilon_{X,p}}{\partial t} &= i \left[\omega_p \epsilon_{X,p} + \lambda_p \epsilon_{X,p} - \frac{\mu}{T_{\rm cm}^3} \int dr' r'^2(s_{r'} \epsilon_{X,p} - s_p \epsilon_{Y,r'} - \bar{s}_{r'} \epsilon_{X,p} + s_p \bar{\epsilon}_{Y,r'})\right]\\
  \frac{\partial \bar{\epsilon}_{X,p}}{\partial t} &= i \left[-\omega_p \bar{\epsilon}_{X,p} - \lambda_p \bar{\epsilon}_{X,p} - \frac{\mu}{T_{\rm cm}^3} \int dr' r'^2(s_{r'} \bar{\epsilon}_{X,p} - \bar{s}_{p} \epsilon_{Y,r'} - \bar{s}_{r'} \bar{\epsilon}_{X,p} + \bar{s}_{p} \bar{\epsilon}_{Y,r'})\right]
\end{align}
with $Y=L$ for $X=R$ and $Y=R$ for $X=L$ and the momentum dependence indicated by subscripts.

In the regime where collective solutions exist, the off-diagonal terms have the following form:
\begin{align}
 \epsilon_{X,p}(t) = Q_{X,p} \exp(-i\Omega t)\ , \qquad \bar{\epsilon}_{X,p}(t) = \bar{Q}_{X,p} \exp(-i\Omega t)\ ,
\end{align}
for $X=R,L$. If $\kappa = \mathrm{Im}(\Omega) > 0$, the solution exhibits exponential growth. 
With this ansatz, we write the equations as
\begin{align}
  \left[\Omega + \omega_p + \lambda_p - \mu \int \frac{dr'r'^2}{T_{\rm cm}^3}(s_{r'}  - \bar{s}_{r'} )\right] Q_{X,p} = \mu s_p \int dr' r'^2(- Q_{Y,r'} + \bar{Q}_{Y,r'})\ ,\\
  \left[\Omega - \omega_p - \lambda_p - \mu \int dr' r'^2(s_{r'} - \bar{s}_{r'} )\right] \bar{Q}_{X,p}  = \mu  \bar{s}_{p}\int dr' r'^2(- Q_{Y,r'} + \bar{Q}_{Y,r'})\ .
\end{align}
Since the integrals and $\mu$ are independent of $p$, we can now write:
\begin{align}
  Q_{X,p} = A_X \frac{s_p}{\Omega + \omega_p + \lambda_p - \mu \int dr' r'^2(s_{r'} - \bar{s}_{r'})}\ ,\\
  \bar{Q}_{X,p} = \bar{A}_X \frac{\bar{s}_{p}}{\Omega - \omega_p - \lambda_p - \mu \int dr' r'^2(s_{r'} - \bar{s}_{r'})}\ ,
\end{align}
where $A_X$ and $\bar{A}_X$ are yet to be determined constants.
These have to fulfil the relations
\begin{equation}
  \label{eq:AX}
  A_X = ( \bar{I} \bar{A}_Y - I A_Y)\ , \qquad \bar{A}_X = ( \bar{I} \bar{A}_Y - I A_Y)\ ,
\end{equation}
where $Y=L$ for $X=R$, $Y=R$ for $X=L$, and
\begin{align}
  I &= \int dr r^2\frac{\mu s_r}{\Omega + \omega_{r} + \lambda_{r} - \mu \int dr' r'^2 (s_{r'} - \bar{s}_{r'})}\ ,\label{eq:Idef}\\
  \bar{I} &= \int dr  r^2\frac{\mu \bar{s}_{r}}{\Omega - \omega_r - \lambda_r - \mu \int dr' r'^2(s_{r'} - \bar{s}_{r'})}\ .\label{eq:Ibardef}
\end{align}
From \eref{eq:AX}, it is clear that $A_X = \bar{A}_X$, and it is possible to find $\Omega$ by numerically solving the following equation: 
\begin{equation}
  \label{eq:Irelation}
  1 = (\bar{I}-I)^2\ .
\end{equation}

\begin{figure}[tbp]
 \centering
 \includegraphics[width=\textwidth]{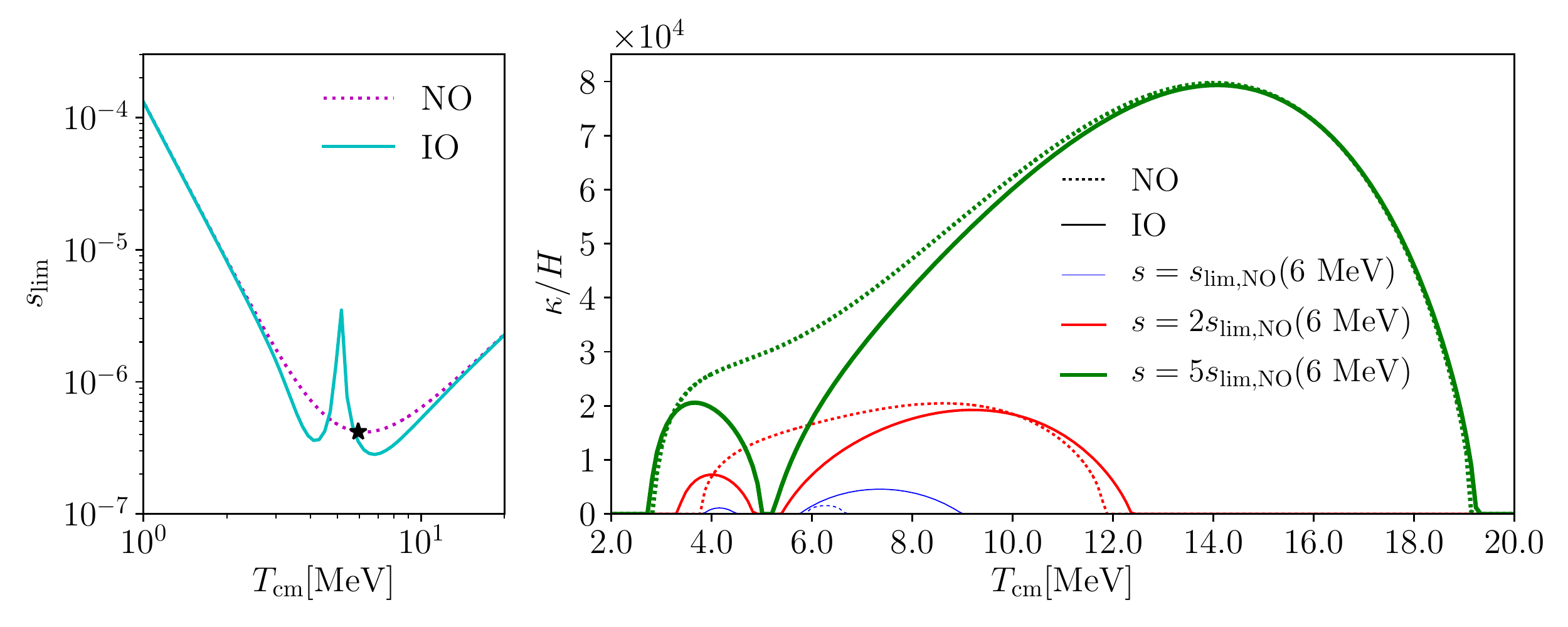}\\
 \caption{{\it Left panel}: Threshold for flavor stability as a function of the temperature. The threshold increases at large temperatures due to the matter term and at low temperatures due to the vacuum term. {\it Right panel}: Growth rate relative to the Hubble expansion rate as a function of the temperature for different values of $s$. We assume $\bar{s} = s$ and use $s_{\rm lim} = 4.2\times 10^{-7}$ corresponding to $T_{\rm cm}=6$~MeV for NO as a reference value. The growing instability is symmetric in $L \leftrightarrow R$ (isotropic) for IO below $\sim 5\;{\rm MeV}$ and antisymmetric (anisotropic) for IO above $\sim 5\;{\rm MeV}$ and for NO. In both panels we use $|\Delta m^2| = 4.5 \times 10^{-3}~{\rm eV}^2$. The star denotes $s_{\rm lim,NO}(6\;{\rm MeV})$.}
 \label{fig:growthrate_slim}
\end{figure}

We are interested in determining the values of $s$ where the collective instability is present. 
Since the system we consider is close to equilibrium and has an approximate symmetry between neutrinos and antineutrinos, we assume that $s_p = \bar{s}_p = s f_0(p)$, where $f_0(p) = 1/[\exp(p/T_{\rm cm}) +1]$ is a Fermi-Dirac distribution. With this assumption, the term proportional to $\mu$ in the denominator vanishes, and we can evaluate the integrals numerically.
The limit of stability is found for ${\rm Re}(\Omega)=0$ and ${\rm Im}(\Omega) \rightarrow 0$. Hence we determine $s_{\rm lim}$ by setting $\Omega/\mu = (10^{-10} i)$ and solving \eref{eq:Irelation} for $s$.
The result is shown in the left panel of \fref{fig:growthrate_slim}. At $T_{\rm cm} \sim 5\;{\rm MeV}$, the limit goes to large values for IO. This is caused by the cancellation of the matter and vacuum term. In NO, such a cancellation is not present and the transition is smooth. At high temperatures, the matter term dominates, and the limit scales as $\lambda/\mu \propto T_{\rm cm}^2$, while at low temperatures the vacuum term dominates and the limit scales as $\omega/\mu \propto T_{\rm cm}^{-4}$.

The asymptotic scaling can also be recovered by using a single energy approximation. The integrals can be approximated by
\begin{equation}
  I_S = \frac{\mu s}{\Omega + \omega_S + \lambda_S} \frac{3 \zeta(3)}{2}\ , \qquad
  \bar{I}_S = \frac{\mu s}{\Omega - \omega_S - \lambda_S} \frac{3 \zeta(3)}{2}\ ,
\end{equation}
where
\begin{equation}
  \label{eq:omegaSlambdaS}
  \omega_S = \frac{\Delta m^2}{2\left<p\right>}\;, \qquad \lambda_S = \frac{14\sqrt{2}G_{\rm F} \pi^2 T_{\rm cm}^4}{45 m_W^2 \left<1/p\right>} \;,
\end{equation}
with $\left<p\right> \approx 3.15 T$ and $(\left<1/p\right>)^{-1} \approx 2.19 T$. In this approximation, \eref{eq:Irelation} can be solved directly for $\Omega$:
\begin{equation}
  \label{eq:OmegaS}
  \Omega_S = \pm \sqrt{ (\omega_S + \lambda_S)^2 \pm 2 \mu_S s (\omega_S + \lambda_S)}\ ,
\end{equation}
with $\mu_S = \mu\; 3\zeta(3)/2$. For an imaginary solution to exist, we need
\begin{equation}
  \label{eq:slimS}
  2 \mu_S s > \omega_S + \lambda_S \;.
\end{equation}
This simple condition can be used to estimate where we expect collective oscillations in our numerical results.

To fulfil \eref{eq:Irelation}, the integrals must satisfy
\begin{equation}
  \label{eq:Irelation2}
  \bar{I} - I = \pm 1\ .
\end{equation}
If the difference is positive, \eref{eq:AX} implies that $A_R = A_L$; if it is negative, we obtain $A_R = -A_L$. These two cases correspond to conserving ($+$) and breaking ($-$) an approximate symmetry between right and left moving neutrinos and hence also to isotropic and anisotropic flavor conversion.

Evaluating the full integrals and assuming $s>0$, we find $\bar{I}-I<0$ for NO ($\omega>0$). This indicates anisotropic flavor conversion. For IO ($\omega<0$) and at high temperatures, we also find $\bar{I}-I<0$. However, at $T<5\;{\rm MeV}$, we find the opposite sign corresponding to isotropic flavor conversions. This is in agreement with the results in two dimensions of Ref.~\cite{Cirigliano:2017hmk} and  in line with our comparison to Ref.~\cite{Raffelt:2013rqa}. Notice that if $s<0$ then the sign of $I$ and $\bar{I}$ changes, and the role of the two orderings is reversed.
As the asymmetry between $\nu_e$ and $\nu_x$ rises, due to heating from electron-positron annihilation, the collective solutions are triggered when $s=\rho_{ee}-\rho_{xx}$ crosses $s_{\rm lim}$.
  The off-diagonal part of the density matrix grows exponentially with the growth rate $\kappa = {\rm Im}(\Omega)$ shown in the right panel of \fref{fig:growthrate_slim}. The growth rate is determined by solving \eref{eq:Irelation} for $\Omega$, given the values of $s$ and $T_{\rm cm}$. 
  We show the growth rate normalized to the Hubble expansion rate as a function of the temperature for a few different values of $s$. As in the left panel, the transition for IO from anisotropic instabilities at high temperatures to isotropic instabilities at low temperatures is seen at $\sim 5\;{\rm MeV}$. In NO, the instabilities are anisotropic in nature at all temperatures where they are present.
As the scale of the $y$-axis demonstrates, the growth rates are much larger than the Hubble expansion rate, and the growth rate increases as $s$ becomes increasingly larger than $s_{\rm lim}$. Thus, we expect any tiny perturbation to grow to non-linear scales when the instability is triggered. If $s$ is gradually created, e.g., by  heating from electron-positron annihilations, we expect the solution to enter the non-linear regime shortly after $s$ has crossed $s_{\rm lim}$.

\section{Numerical solutions: isotropic setup}
\label{sec:isotropic}
 Before we present the results for the antisymmetric modes identified in the linear stability analysis that allow for the breaking of the approximate isotropy, we review the case of isotropic conversions and investigate how the different parts of the collision term affect the results. In this section, after introducing our numerical setup, we describe the flavor conversion phenomenology for our two beam model with perfectly isotropic initial conditions. This is equivalent to the assumption of isotropy considered elsewhere in the literature. In addition, we explore the effect on the flavor conversion physics of the different terms entering the collision term.

\subsection{Numerical framework}
For the numerical implementation of the full set of non-linear QKEs, we express the matrices of densities as polarization vectors:
\begin{equation}
 \label{eq:poldef}
 \rho_X(r) r^2 = \frac{1}{2} [P_{0,X}(r) + \mathbf{P}_X(r) \cdot \sigma]\ , \quad \bar{\rho}_X(r) r^2 = \frac{1}{2} [\bar{P}_{0,X}(r) + \bar{\mathbf{P}}_X(r) \cdot \sigma]\ ,
\end{equation}
where $\sigma$ is a vector of Pauli matrices  and $r=p a$ is the comoving momentum for momentum $p$ and scale factor $a$, $X \in \{R,L\}$, $\mathbf{P}_X = (P_{x,X}, P_{y,X}, P_{z,X})^T$, and $\bar{\mathbf{P}}_X = (\bar{P}_{x,X}, \bar{P}_{y,X}, \bar{P}_{z,X})^T$.
In terms of the polarization vectors, the QKEs can be rewritten as real valued equations. We use a modified version of LASAGNA~\cite{Hannestad:2013pha,Hannestad:2012ky} to solve the equations (see Appendix~\ref{app:QKE} for details).

The mixing parameters used throughout the paper are~\cite{Esteban:2020cvm}\footnote{See also \href{www.nu-fit.org}{NuFIT 5.0 (2020), www.nu-fit.org}.}
\begin{equation}
 \label{eq:mixingNO}
\textrm{NO:}\quad \Delta m^2 = 2.51 \times 10^{-3}{\rm eV}^2, \quad \sin^22\theta = 0.0888, 
\end{equation}
and
\begin{equation}
 \label{eq:mixingIO}
\textrm{IO:}\quad \Delta m^2 = -2.50 \times 10^{-3}{\rm eV}^2, \quad \sin^22\theta = 0.0896.
\end{equation}
We start at a temperature of $40$~MeV
and use 200 comoving momentum bins from $r=0.1$ to $r=20$. The equations are solved using a Runge-Kutta-Domand-Prince solver with relative and absolute tolerances set to $10^{-11}$.

At low temperatures, below $\sim 1$~MeV, vacuum conversions dominate and following them numerically becomes very demanding computationally. However, their behavior is analytically known; by going to a rotating frame, we can exclude them from the numerical solution (see Appendix~\ref{app:QKE} for details).
 In spite of that, at $T_\gamma<0.4$~MeV, the code proceeds very slowly. To further speed it up, we neglect the neutrino-neutrino forward scattering below $T_\gamma \lesssim 0.4$~MeV.

\subsection{Neutrino flavor mixing with two angle bins}

\begin{figure}[b]
 \centering
 \includegraphics[width=\textwidth]{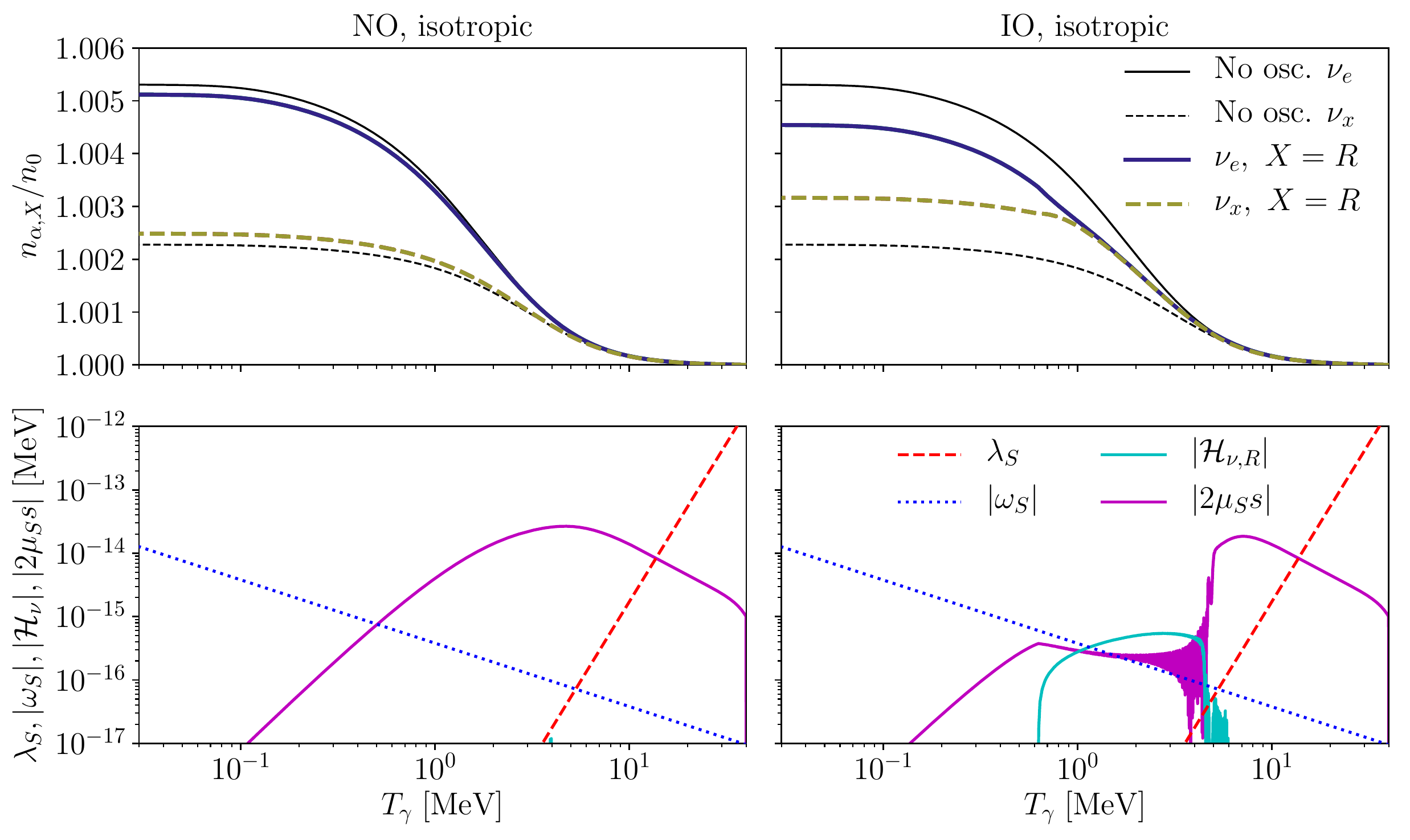}
 \caption{Flavor conversions in our model with two angle bins for perfectly isotropic initial conditions in NO (left) and IO (right). {\it Top panels}: Evolution of $n_{\alpha,X}/n_0$. The black lines represent the case with no flavor conversions. Because of the isotropic conditions, all neutrinos and antineutrinos of a given flavor give identical results. {\it Bottom panels}: The matter-, vacuum-, and neutrino-neutrino potentials as well as $2 \mu_S s$ [see Eqs.~(\ref{eq:Hnu}) and (\ref{eq:omegaSlambdaS})]. 
 }
 \label{fig:isosym}
\end{figure}

Figure~\ref{fig:isosym} shows the results for perfectly isotropic initial conditions in NO (IO) on the left (right). The black lines show the case of no conversions in both top panels. The lower panels show the potentials as well as $2\mu_S s$ which can be compared directly to $\lambda_S$ and $\omega_S$ through the linear stability condition in \eref{eq:slimS}. The general increase in the relative density as the temperature decreases is due to heating from electron-positron annihilations. The case of flavor conversions in NO (left panels) shows little flavor conversion. In fact, as visible from the bottom left panel, $2 \mu_S s$ evolves smoothly without encountering any instability which is what is expected from the linear stability analysis where the symmetric (and therefore isotropy conserving modes) are stable. This is also reflected in the fact that the neutrino-neutrino term, $\mathcal{H}_{\nu\nu}$ is negligible at all temperatures. The small amount of flavor conversion that is seen in the upper left panel is due to vacuum mixing.

 The IO case is shown in the right panels. The upper panel shows significantly more flavor conversion than the NO case. This is consistent with the linear stability analysis where the symmetric modes, which can evolve even in an isotropic system, become unstable below $\sim 5$~MeV. The lower panel clearly demonstrates how this instability is encountered. Initially, when the condition in \eref{eq:slimS} is verified, as $2 \mu_S s$ crosses $\lambda_S$, nothing significant happens, since only the antisymmetric solution becomes unstable. However, as the temperature reaches $\sim 5$~MeV and $\lambda_S \sim \omega_S$, the symmetric, isotropy conserving solution becomes unstable and flavor conversions set in.
 The neutrino-neutrino potential starts to rise at temperatures slightly above $5$~MeV, but efficient conversions only start when the full range of momentum modes becomes unstable.
Once the flavor instability has been triggered, we see how conversions proceed until the temperature has decreased below $1$~MeV, when the density of neutrinos eventually becomes too low and vacuum conversions take over.
 Flavor conversions keep $\nu_e$ and $\nu_x$ in equilibrium until $\sim 1$~MeV, and only neutrinos produced at lower temperatures contribute to the final difference in abundance.

\begin{figure}[tbp]
 \centering
 \includegraphics[width=\textwidth]{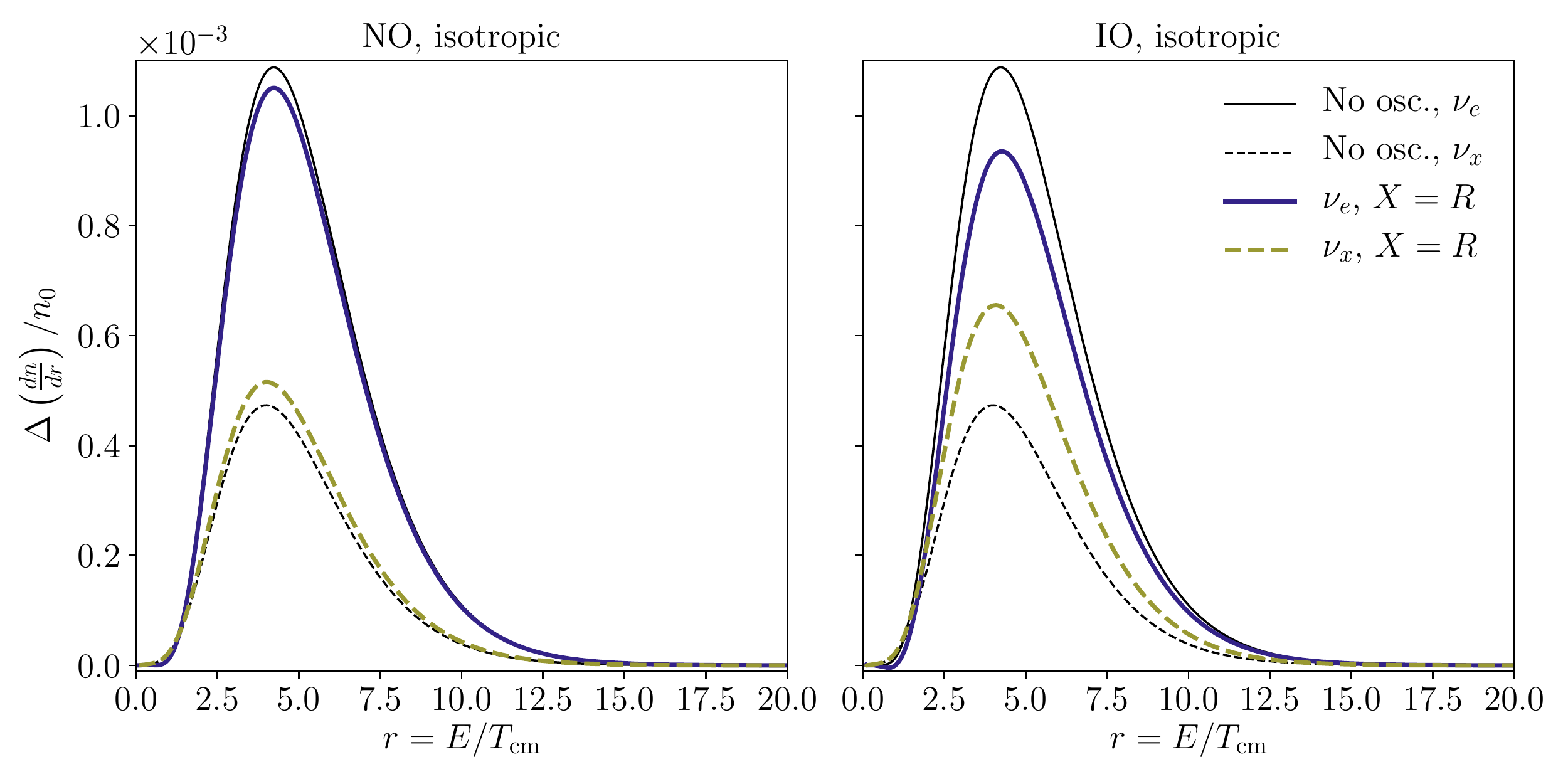}
 \caption{Absolute change in the number distributions of $\nu_e$ and $\nu_x$, see \eref{eq:Deltanabs}, at $T_\gamma = 0.01$~MeV for perfectly isotropic initial conditions in NO (left) and IO (right). Since the system is isotropic, the distributions for antineutrinos and left moving neutrinos are coinciding with the right moving ones.}
 \label{fig:final_spectra}
\end{figure}

The absolute change in the number distributions in the presence of flavor conversions for perfectly isotropic initial conditions and for no conversions are shown in \fref{fig:final_spectra}.
 In both mass orderings (NO on the left and IO on the right), the spectra are rather featureless and maintain the same shape that they had in the absence of flavor conversions.
 In the case of NO, the effect of conversions is small, and conversions are damped by incoherent collisions. For IO, conversions tend to equilibrate the spectra of $\nu_e$ and $\nu_x$ until the epoch of weak decoupling, when they cease to be efficient. At lower temperatures, the neutrino spectra continue to be heated, and neither of these processes are expected to give rise to spectral features in \fref{fig:final_spectra}.

\subsection{Dependence on the collision term and forward scattering potentials}

\begin{figure}[tbp]
 \centering
 \includegraphics[width=\textwidth]{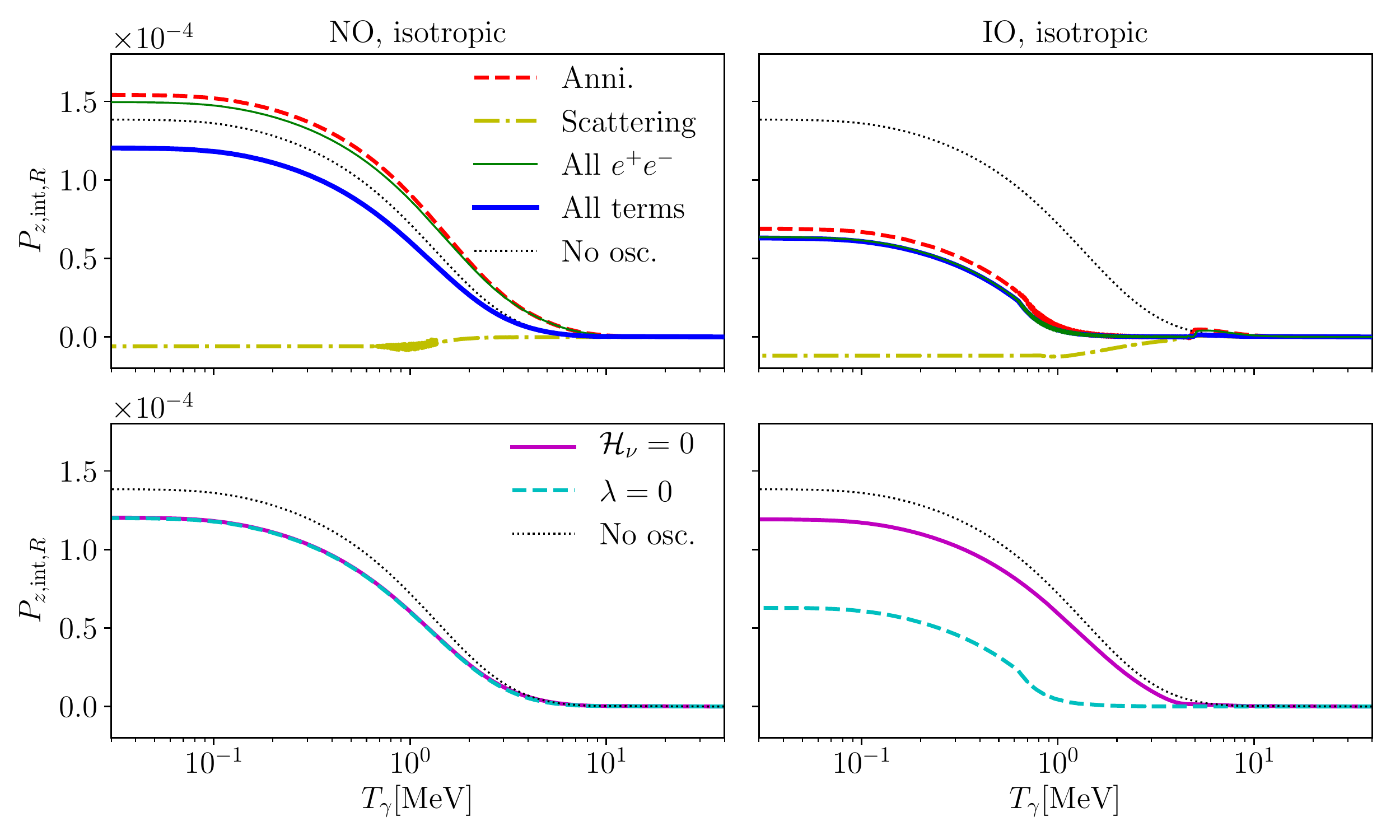}
 \caption{Flavor conversions in our model with two angle bins for perfectly isotropic initial conditions in NO (left) and IO (right) for different contributions to the collision term and Hamiltonian. The top panels include all terms in the Hamiltonian [see Eqs.~(\ref{eq:Hvac})-(\ref{eq:Hnu})] and the collision terms indicated in the legend. The magenta line in the bottom panels is obtained for the neutrino potential $\mathcal{H}_{\nu\nu} =0$ which is an assumption commonly adopted in the literature, see e.g.~Refs.~\cite{Mangano:2005cc,deSalas:2016ztq,Froustey:2020mcq, Bennett:2020zkv}. The dashed cyan line in the bottom panels represents the case with no matter potential: $\lambda =0$. The thin dotted line uses all collision terms but no conversions. The evolution for left moving neutrinos as well as for all antineutrinos is identical to the one of right moving neutrinos for all of these runs. The magenta and cyan curves are on top of each other in the lower left panel, and the thick blue and thin green curves are on top of each other in the upper right panel.}
 \label{fig:collpottest}
\end{figure}

In the following, we will focus on the effect of the various contributions to the collision term on flavor conversions. To this purpose, we compare the $z$-component of the polarization vectors.
The integrated $z$-component of the polarization vector is~\footnote{Note that $P_{z,X}$ is defined to have a factor of $r^2$ with respect to $\rho$.}
\begin{equation}
 \label{eq:Pzint}
 \Pz{X} = \int \frac{dr}{4\pi^2} P_{z,X}(r) \qquad {\rm for} \quad X \in \{R,L\} \; .
\end{equation}
The evolution of $\Pz{X}$ for four different cases of the collision term is shown in the top panels of \fref{fig:collpottest} for NO and IO.
 The first case is ``Annihilation'', where only simple annihilations between neutrinos and electron/positron pairs are included in the diagonal terms. For $\mathcal{C}_{ex}$, the simple damping term is included (we neglect coherent annihilations by setting $y_G=0$). For ``Scattering'', it is similarly only scatterings between neutrinos and electrons or positrons that are included in $\mathcal{C}_{\alpha\alpha}$, and the damping term is used for $\mathcal{C}_{ex}$. In the case of ``All $e^+e^-$'', all interactions with electrons and positrons are included in the collision term. Finally, for ``All terms'' we also include the neutrino-neutrino interactions and use the full approximation for the collision term.
 The coefficients corresponding to the four cases are listed in \tref{tab:colltest}.

\begin{table}
 \caption{$y_x$ for the different constants in \eref{eq:Gammax} for the cases seen in \fref{fig:collpottest}. $y_G$ and $y_D$ have identical values. The corresponding reactions are listed in \tref{tab:collisions}.
 }
 \centering
 \begin{tabular}{l c c c c c c c c c}
 \hline\hline
 Case & $y_{s,e}$ & $y_{s,x}$ & $y_{a,e}$ & $y_{a,x}$ & $y_D$ & $y_{G}, y_{C} $ & $y_{d}$ & $y_{\nu}$ & $y_{\bar{\nu}}$\\
 \hline
 Annihilation & 0 & 0 & 0.049 & 0.010 & 0.296 & 0 & 0 & 0 & 0 \\
 Scattering & 0.196 & 0.041 & 0 & 0 & 0.296 & 0 & 0 & 0 & 0 \\
 All $e^+e^-$ & 0.196 & 0.041 &0.049 & 0.010 & 0.296 & -0.006 & -0.044 & 0 & 0 \\
 All terms & 0.196 & 0.041 &0.049 & 0.010 & 0.296 & -0.006 & -0.044 & 0.061 & 0.0206\\
 \hline\hline
 \end{tabular}
 \label{tab:colltest}
\end{table}

For NO in the upper left panel, there are no conversions except for the case where only scatterings are included.
The no-conversion result with all collision terms is plotted with a dotted line for reference; both for annihilations into $e^+e^-$ and for all $e^+e^-$ terms, the results show a slight increase with respect to the case without flavor conversions. This is explained by the lack of neutrino-neutrino interactions which tend to equilibrate the two flavors and thereby reduce $\Pz{X}$.
When only scatterings with $e^+e^-$ are included, the collision term does not change the number of neutrinos and antineutrinos and, hence, we do not see the increase in $\Pz{X}$ that is present in the other cases. However, the elastic scattering still allows for a transfer of energy, therefore the momentum distributions are distorted (see also Appendix~\ref{app:collterm}). Consequently, the electron flavors dominate at high momentum while the non-electron flavors dominate at low momentum. Since conversions and collisions are most efficient at converting neutrinos at high momentum, the larger densities of $\nu_x$ and $\bar{\nu}_x$ at low momentum lead to a slight decrease in $\Pz{X}$ as the dot-dashed yellow curve shows.
As $s \propto \Pz{X}$ is now negative, we know from the stability analysis that the isotropic mode is unstable for NO. Indeed this is confirmed by the numerical results, where conversions occur below $\approx 1.5$~MeV.
The calculation with all collision terms (thick blue curve) shows no clear oscillations in NO. This result corresponds to the left panels of \fref{fig:isosym} and \fref{fig:final_spectra}.

For IO, the situation is reversed, as shown in the upper right panel of \fref{fig:collpottest}. In the cases where annihilations are included (the dashed red, the thin green, and the thick blue curves), the flavor evolution as a function of the temperature is almost identical in all cases, and we see a clear suppression of $\Pz{X}$ due to conversions introduced by the isotropic instability which is triggered at $\sim 5$~MeV. This is in agreement with the results from the linear stability analysis since $s$ is now positive. Initially, flavor conversions are efficient enough to keep $\Pz{X}$ close to zero, but as the temperature decreases and the neutrino-neutrino forward scattering term becomes subdominant, $\Pz{X}$ starts to increase and reaches approximately half of the no-oscillation value.
As in NO, the dot-dashed yellow curve with scatterings on $e^+e^-$ shows a slight decrease of $\Pz{X}$, but here without any oscillation since both $s$ and $\omega$ are negative.

The Hamiltonian that governs the flavor evolution receives contributions from vacuum, matter and neutrinos terms that are described in Eqs.~(\ref{eq:Hvac})-(\ref{eq:Hnu}). In the bottom panels of \fref{fig:collpottest}, the effects of matter and neutrino potentials are shown. For the magenta curve, the neutrino-neutrino forward scattering is neglected. The results are indistinguishable from the thick blue curve in the upper left panel, the reason being that no collective neutrino instabilities are encountered; hence, the neutrino-neutrino forward scattering does not have any large effects.
Similarly, we find that the dashed blue curve in the lower left panel, where the matter potential is neglected, is identical to the thick blue curve in the upper left panel. Some differences can be expected at temperatures above $5$~MeV, where the matter potential dominates the vacuum potential, but at these temperatures, the deviation from equilibrium is too small to be visible on the scale displayed in the plot. Also, it has no impact on the final flavor content at low temperature.

For IO, the results are shown in the lower right panel. Here, the neglected neutrino-neutrino potential for the magenta curve has the very prominent effect of removing the collective instability. This is the assumption used in~\cite{Mangano:2005cc,deSalas:2016ztq,Froustey:2020mcq, Bennett:2020zkv}. Hence, $\Pz{X}$ is twice as large at low temperatures as it is for the thin green curve in the lower left panel. For the blue dashed curve, where the neutrino-neutrino potential is included, the collective instability is still present, and the final value of $\Pz{X}$ is very close to the value found for the thick blue curve in the upper right panel.

To summarize, our calculations with isotropic initial conditions confirm what we expected from the linear stability analysis (see Sec.~\ref{sec:linear}). The NO case shows a very small effect due to conversions, while the difference between $\nu_e$ and $\nu_x$ (quantified by the magnitude of $P_z$) is cut in half by conversions for IO. 
For anisotropic initial conditions, we expect NO to show conversions as well, since the antisymmetric modes will be allowed to grow.

\section{Numerical solutions: anisotropic setup}
\label{sec:anisotropic}
In this section, we 
explicitly introduce a small symmetry breaking between right and left moving neutrinos to explore the neutrino flavor conversion physics in an anisotropic setup.
In order to do that, we define the initial value of $P_x$ as follows
\begin{align}
 \label{eq:Pxani}
 P_{x,R} &= B_i r^2 f_0 ,\\
 P_{x,L} &= -B_i B_L r^2 f_0 ,
\end{align}
where $f_0 = 1/(\exp(r)+1)$, $P_{y,R} = P_{y,L} =0$, and the anisotropy of the initial conditions depends on the parameters $B_i$ and $B_L$.
 The same initial conditions are used for antineutrinos for $\bar{P}_x$ and $\bar{P}_y$, while the other components of $\mathbf{P}$ and $\mathbf{\bar{P}}$ are defined as in the isotropic case. In the following, we explore the flavor conversion phenomenology in NO and IO within this framework.

\subsection{Normal mass ordering}

\begin{figure}[tbp]
 \centering
 \includegraphics[width=\textwidth]{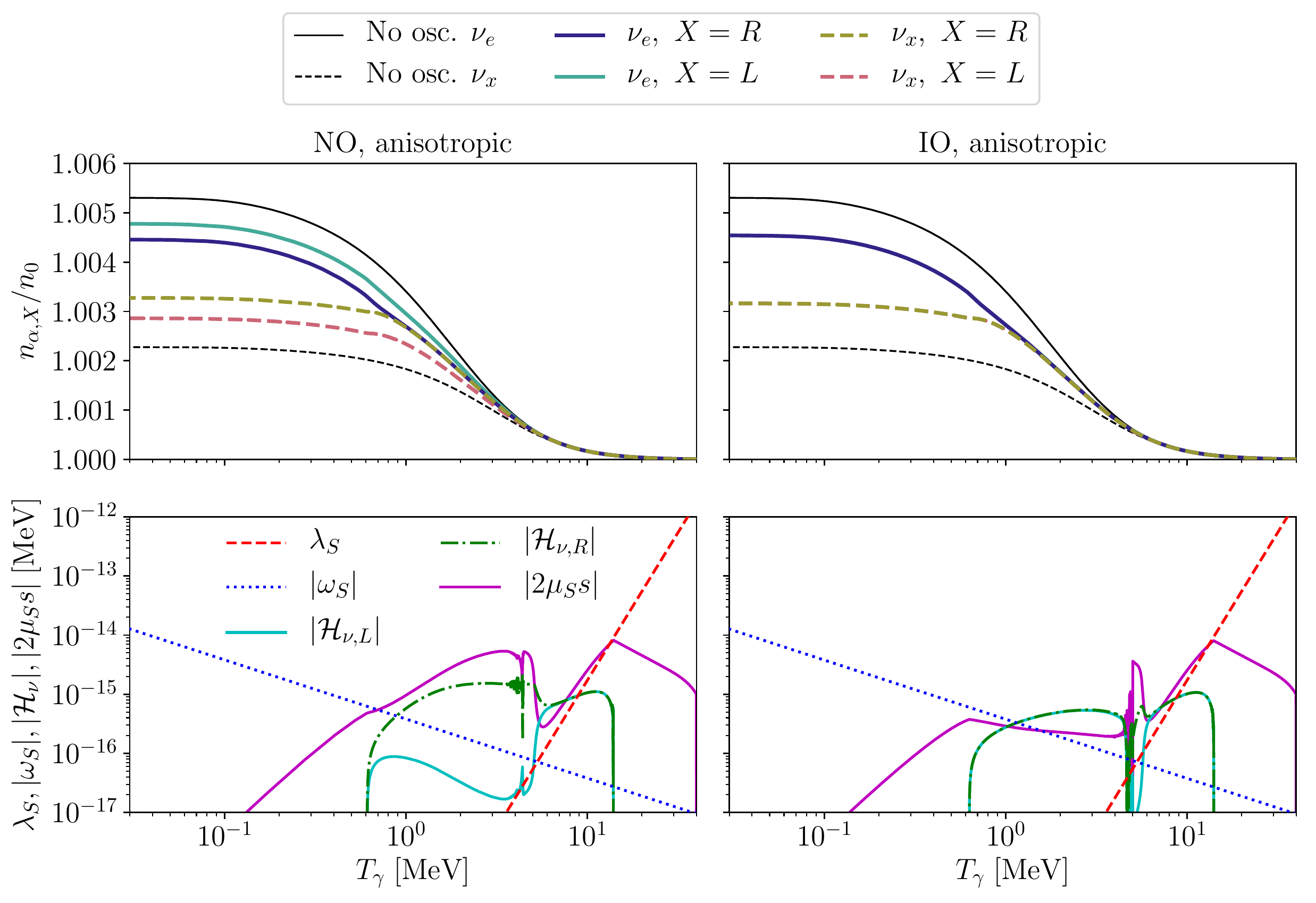}
 \caption{Flavor conversions in our model with two angle bins for anisotropic initial conditions in NO (left) and IO (right). {\it Top panels}: Evolutions of $n_{\alpha,X}/n_0$ for left and right moving neutrinos as functions of the temperature. The black lines refer to the case with no conversions and, hence, identical evolution for neutrinos and antineutrinos moving in both directions. In the upper left panel, all neutrinos and antineutrinos of a given flavor give identical results. In the upper right panel, neutrinos and antineutrinos of a given flavor and moving in the same direction are on top of each other. {\it Bottom panels}: Evolution of the matter-, vacuum-, and neutrino-neutrino potentials as well as $2 \mu_S s$ [see \eref{eq:Hnu} and (\ref{eq:omegaSlambdaS})] as functions of the temperature. In NO, the asymmetry between left and right moving modes is amplified by flavor conversions, while results similar to the isotropic case are found for IO. 
}
 \label{fig:anisym}
\end{figure}

\begin{figure}[tbp]
 \centering
 \includegraphics[width=\textwidth]{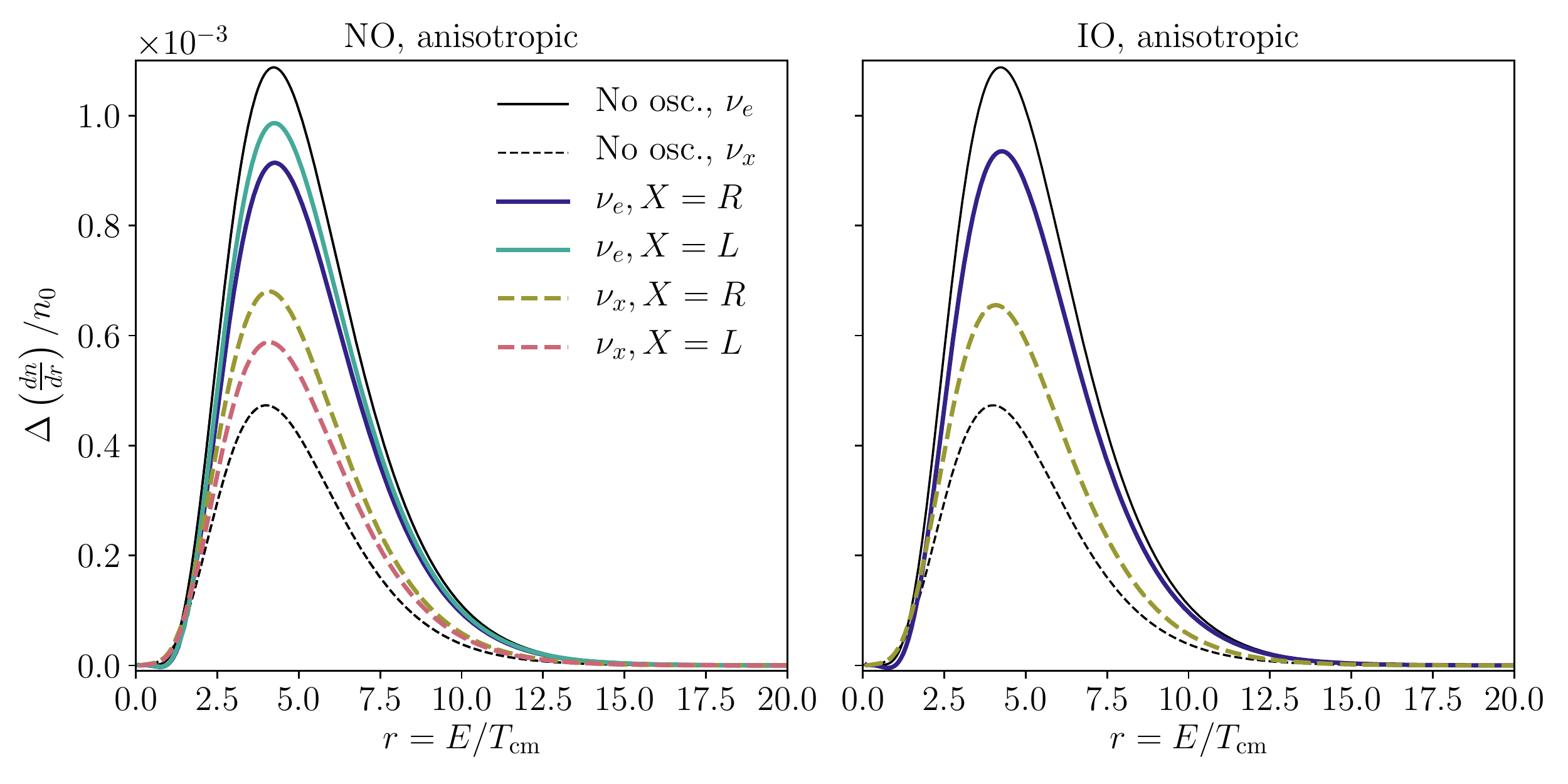}
 \caption{Absolute change in the number distributions, see \eref{eq:Deltanabs}, at $T_\gamma = 0.01$~MeV for $\nu_e$ and $\nu_x$ for anisotropic initial conditions in NO (left) and IO (right). The distributions for right and left moving neutrinos are coinciding in the right panel. All antineutrinos have distributions identical to the corresponding neutrinos.}
 \label{fig:final_spectra_ani}
\end{figure}

From the linear stability analysis, we expect to see a significant difference between the results with isotropic and anisotropic initial conditions in NO (see Fig.~\ref{fig:growthrate_slim}). Figure~\ref{fig:anisym} shows the corresponding numerical solution of the QKEs for $B_i = 10^{-15}$ and $B_L = 1$.
 At high temperature, the solutions do not strongly depend on the mass ordering, as we will also discuss in the next section. The lower left panel of \fref{fig:anisym} shows that $2\mu_S s$ starts below $\lambda_S$ and gradually increases until it reaches $\lambda_S$ at $T_\gamma \simeq 14$~MeV. At this point, the antisymmetric instability identified in Fig.~\ref{fig:growthrate_slim} kicks in. As a consequence, $\mathcal{H}_{\nu\nu}$ grows quickly followed by an increase in the difference between $|\mathcal{H}_{\nu\nu,R}|$ and $|\mathcal{H}_{\nu\nu,L}|$. 
At $\simeq 6$~MeV, the relative difference between $|\mathcal{H}_{\nu\nu,R}|$ and $|\mathcal{H}_{\nu\nu,L}|$ 
reaches order unity, and the non-linear evolution sets in. 
In this case, a large difference between $|\mathcal{H}_{\nu\nu,R}|$ and $|\mathcal{H}_{\nu\nu,L}|$ develops and persist until sub-MeV temperatures. The transition from matter to vacuum dominated regimes at $\simeq 5$~MeV leads to rapid oscillations in $\mathcal{H}_{\nu\nu, X}$ and $2\mu_S s$, but the overall behavior is the same above and below $5$~MeV.

We stress that, in this anisotropic setup, the asymmetry between left and right moving modes in NO is amplified by flavor conversions and it persists even at low temperatures.
As we assume homogeneity in our model, the amount of flavor conversion is the same for every position in space.
In the upper left panel of \fref{fig:anisym}, the right moving neutrinos are closer to flavor equilibrium than the left moving neutrinos. However, we have proven (results not shown here) that the $L/R$ ordering is sensitive to the chosen initial conditions, although the qualitative features are not. 
We find that different values of $B_L$ do not change the $L/R$ ordering, while the ordering swaps with the sign of $B_i$. 

The absolute change in the number distributions is shown in \fref{fig:final_spectra_ani} where the NO case is displayed in the left panel. Comparing the NO anisotropic results to the isotropic results in the left panel of \fref{fig:final_spectra}, we notice the significantly larger conversion probability, and that the left and right moving modes have separated. This is in good agreement with the results in \fref{fig:anisym}, and shows that there are no particular features present in the final number distribution.

\begin{figure}
  \centering
  \includegraphics[width=0.75\textwidth]{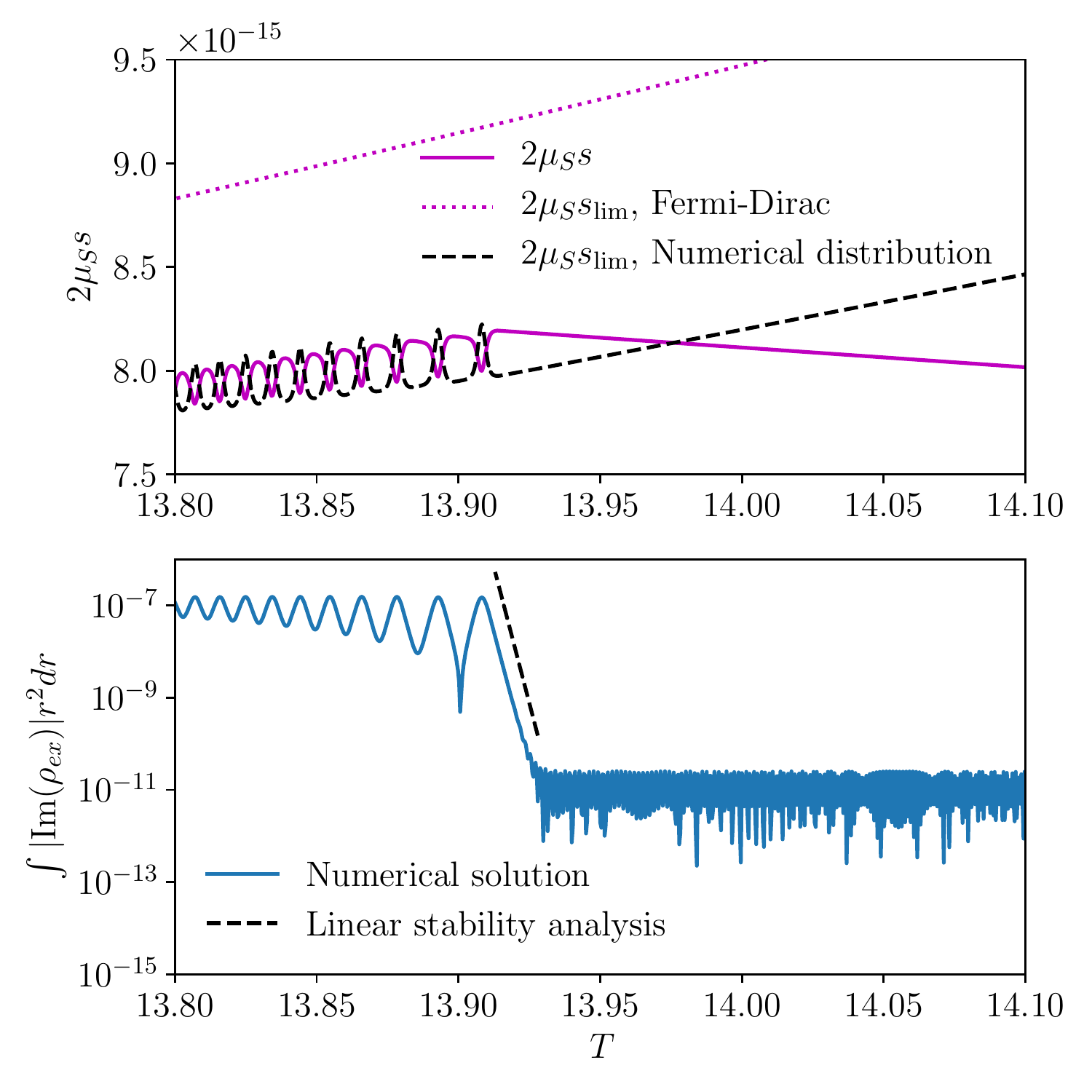}
  \caption{ Comparison of numerical results and linear stability analysis. {\it Upper panel}: $s=\rho_{ee}-\rho_{xx}$ times $2\mu_S$ as in the lower left panel of \fref{fig:anisym}. The results of the numerical solution (magenta) is shown as well as limits from linear stability analyses using a Fermi-Dirac distribution for $s_p$ (magenta dotted) and using the distribution for $s_p$ from the numerical solution (black dashed).
  {\it Lower panel}: The imaginary part of the density matrix integrated over momentum. The black dashed line indicates the growth rate determined through the linear stability analysis at $T=13.92\;{\rm MeV}$ using the numerical distribution. The vertical off-set is arbitrary and chosen for easy comparison.}
  \label{fig:lin_growth}
\end{figure}

As a consistency check, we consider the temperatures where the non-linear behavior starts, i.e.~around $14\;{\rm MeV}$ in \fref{fig:lin_growth}. The upper panel shows a zoom on $2\mu_S s$ from the lower left panel of \fref{fig:anisym}. In addition, we show the stability limit for $s$ (magenta dotted line) that was derived in \sref{sec:linear} using a Fermi-Dirac distribution for $s_p$ and shown in the left panel of \fref{fig:growthrate_slim}. However, the linear stability analysis with Fermi-Dirac distributions predicts that the non-linear evolution starts at slightly lower temperatures than what we find in the numerical solution. The culprit turns out to be the description of $s_p$ by a Fermi-Dirac distribution. Instead, we use the distribution of $s_p$ determined in the numerical solution (black dashed line). With this approach, the numerical solution and linear stability analysis are in good agreement since the non-linear evolution starts shortly after the numerical solution crosses the limit of stability.
  In order to quantify the agreement further, we consider the off-diagonal part of the density matrix and its growth. In the lower panel of \fref{fig:lin_growth}, we show the integrated imaginary part which most clearly shows the growth. Below $T=13.93$~MeV, the exponential growth is visible. At slightly higher temperatures, the solution is also predicted to be unstable according to the upper panel, but in that case the growth is hidden by the finite value of $\int |{\rm Im}(\rho_{ex})|r^2 dr$. As a further cross check, we calculated the expected growth rate at $T=13.92\;{\rm MeV}$ using the linear stability analysis which is indicated by the dashed black line in the lower panel; this is  in good agreement with the growth rate seen in the numerical simulations. It should be noted that the offset between the numerical results and linear stability analysis is introduced for legibility. 

\subsection{Inverted mass ordering}
The right panel of \fref{fig:final_spectra_ani} shows the outcome of our numerical simulations in IO. One can see that the final flavor outcome is indistinguishable from the one with isotropic initial conditions shown in \fref{fig:final_spectra}. 
This is also true when considering the evolution of the density matrix elements as functions of the temperature, where the results in the upper right panel of \fref{fig:anisym} correspond to the blue curve in the upper right panel of \fref{fig:isosym}.
Such behavior is explained because only the isotropic mode is unstable below $5$~MeV for IO, and any flavor mixing due to the anisotropic setup above $5$~MeV would have a very small effect on the final flavor outcome, since the heating from electron-positron annihilation does not become efficient until $\simeq 1$~MeV.

The numerical results can be understood in terms of the linear stability analysis.
At high temperatures, $2 \mu_S s$ is smaller than both $\omega_S$ and $\lambda_S$, as shown in the lower right panel of \fref{fig:anisym}. As the lowest energy electrons and positrons become non-relativistic, $s$ grows while the matter potential decreases with falling temperature. At $\simeq 14$~MeV, $2 \mu_S s \simeq \lambda_S$, the antisymmetric instability from the linear stability analysis is encountered, and $2 \mu_S s$ starts to follow the matter potential. As it decreases further with temperature, the system stays as close to stability as possible. This is an effect of the antisymmetric instability, but rather than through conversions, it materializes as a gradual change of the neutrino spectrum such that $2\mu_S s$ can decrease although the neutrinos are still being heated.
At the same time, the difference between $|\mathcal{H}_{\nu\nu,R}|$ and $|\mathcal{H}_{\nu\nu,L}|$ grows, and eventually at a temperature of $\simeq 6$~MeV, the relative difference reaches order unity. A significant difference between right and left moving modes appears as the instability becomes non-linear, but at $T_\gamma\simeq 5$~MeV, the vacuum term starts to dominate the matter term, and the antisymmetric instability disappears while the symmetric instability begins to dominate. Hence, the difference between right and left moving modes quickly decays. Still, the collective effects maintain flavor equilibrium until the temperature reaches $\simeq 1$~MeV, where a difference between $\rho_{ee}$ and $\rho_{xx}$ finally starts to build up. At $T_\gamma\simeq 0.1$~MeV, weak interactions eventually freeze out and neutrino heating ceases.

Until now, we have assumed that neutrinos and antineutrinos have identical abundances, and we have ensured that our numerical solution preserves this symmetry at all temperatures. In the next section, we will relax this assumption and investigate the consequences.

\section{Numerical solutions: anisotropic setup with neutrino-antineutrino symmetry breaking}
\label{sec:symmbreaking}

A very common assumption in calculations of the neutrino flavor evolution in the early universe is the one of perfect symmetry between neutrinos and antineutrinos~\cite{Mangano:2005cc,Ichikawa:2005vw,deSalas:2015glj,deSalas:2016ztq,Gariazzo:2019gyi,Akita:2020szl,Froustey:2020mcq, Bennett:2020zkv}. However, the baryon asymmetry of the universe is $\mathcal{O}(10^{-9})$, which would naturally imply an asymmetry in the lepton sector of the same order, although there is no experimental evidence for that~\cite{Mangano:2011ip,Oldengott:2017tzj}. For a lepton asymmetry of $\mathcal{O}(10^{-9})$, the neutrino self-interaction potential would be extremely small. Although such a small self-interaction potential may seem unlikely to have an effect on the evolution of the neutrino flavor, due to the non-linear nature of $\nu$-$\nu$ interactions, this small asymmetry could be amplified. This is what we find in our simulation results shown in this section.
\begin{figure}[tbp]
 \centering
 \includegraphics[width=\textwidth]{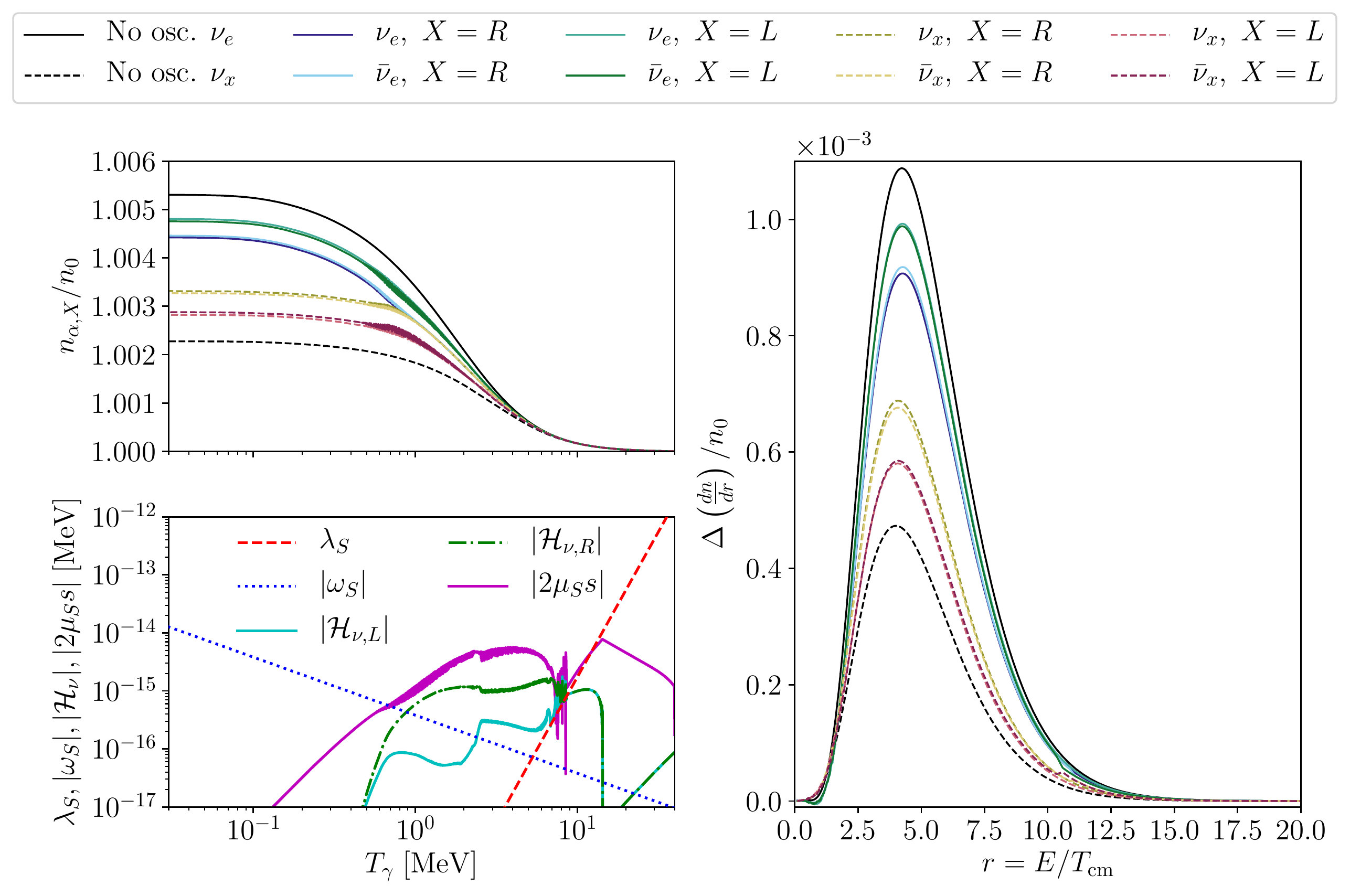}
 \caption{Flavor conversions in our model with two angle bins for anisotropic initial conditions in NO, in the presence of a small lepton asymmetry.
 {\it Top left panel}: Evolution of $n_{\alpha,X}/n_0$ for left and right moving neutrinos and antineutrinos as a function of the temperature. The black lines have no conversions and, hence, identical evolution for neutrinos and antineutrinos moving in both directions. 
 {\it Bottom left panel}: The matter, vacuum, and neutrino-neutrino potentials as well as $2 \mu_S s$ [see Eqs~(\ref{eq:Hnu}) and (\ref{eq:omegaSlambdaS})].
 {\it Right panel}: Absolute change in the number distributions for the cases of no conversions and anisotropic conversions. 
}
 \label{fig:aniasym}
\end{figure}

In order to explore the effects of a small initial asymmetry between neutrinos and antineutrinos, we introduce a small chemical potential: $\mu_{\rm ini} = 10^{-9}$. Our results are shown in \fref{fig:aniasym}; down to $\simeq 10$~MeV, there is no difference between the solutions in Figs.~\ref{fig:anisym} and \ref{fig:aniasym}, but around that temperature, the difference between neutrinos and antineutrinos grows exponentially until it reaches a magnitude where it changes the overall flavor evolution.
The exponential growth persists when the resolution in the co-moving momentum grid is increased and the tolerances of the differential equation solver are tightened.
The total lepton number is conserved to high precision by the collision term, but a small numerical error is accumulated during the evolution. However, the maximal change in the total lepton number is less than $10^{-3}$ of the maximal change in the electron neutrino-antineutrino asymmetry; thus, we conclude that it cannot affect the growth of the asymmetry: the amplification of the neutrino-antineutrino asymmetry is  a physical effect. We stress that this is not a new phenomenon in the context of neutrino-neutrino oscillations, but it has never pointed out in the context of the early universe.

In addition to the numerical checks of our solution, it would be beneficial to have some analytical understanding of this effect. However, the nonlinear evolution in the numerical solution sets in at $T\sim 14\;{\rm MeV}$, and thus it is challenging to describe the growth in the difference between neutrinos and antineutrinos with a linear stability analysis at $T\sim10\;{\rm MeV}$. Instead, we consider a simpler model in Appendix~\ref{app:nunubar} which is well known in the literature~\cite{Hannestad:2006nj}. Here we demonstrate how an exponential growth of lepton asymmetries can arise from initial conditions which show asymmetry between neutrinos and antineutrinos as a part of the so-called bipolar instability. The exponential growth of the asymmetry stems from the  interplay between different terms in the Hamiltonian, and does not involve the collision term.
Although the exponential growth is not affected by the collision term, the subsequent evolution of the neutrino-antineutrino asymmetry is modulated by collisions; the difference between $\nu_e$ and $\nu_x$ is due to the collision term, and the off-diagonal damping term affects oscillations.
The results in \fref{fig:aniasym} refer to the NO case, but a similar growth takes place for IO. However, in that case the final results are almost unchanged when compared to the case with $\mu_{\rm ini} = 0$. For IO the growth of the neutrino-antineutrino asymmetry is also present for isotropic initial conditions, while NO shows no growth in that case. This pattern further demonstrates that the amplification of the neutrino-antineutrino asymmetry is a collective effect since collective oscillations are absent for NO and isotropic initial conditions only.  A more general discussion and analysis of this interesting phenomenon is postponed to future work.

Comparing the left panels in \fref{fig:aniasym} to the left panels in \fref{fig:anisym}, we see a $\simeq 1$\% difference in the heating of neutrinos and antineutrinos in \fref{fig:aniasym} which arise due to the amplification of the initial asymmetry. The breaking of the $L/R$ symmetry again gives two different $L/R$ orderings that are sensitive to the initial conditions. However, we now find that even a tiny change in the value of $\mu_{\rm ini}$ can take the solution from one $L/R$ ordering to the other. Furthermore, there is no simple connection between the values of $B_i$ and $B_L$, where small changes can also modify the $L/R$ ordering. When increasing the momentum resolution and tightening the tolerances, the $L/R$ ordering again changes, and there is no clear convergence. The explanation of this behavior is that the solutions diverge exponentially; this is followed by rapid conversions below $10$~MeV. This trend is a hallmark of chaotic behavior, which is known to be present in collective neutrino conversions~\cite{Hansen:2014paa}, although it has also been demonstrated that seemingly chaotic behavior~\cite{Shi:1996ic, Enqvist:1999zs, DiBari:1999vg, Abazajian:2008dz, Braad:2000zw} can have a numerical origin~\cite{Hannestad:2013pha}. 
Since the choice of tolerances and momentum grid resolution affects how the initial perturbations are amplified, one cannot expect to see detailed convergence of the results below $10$~MeV.
However, runs with higher resolution and tighter tolerances confirm the features shown in \fref{fig:aniasym} and only the $L/R$ ordering remains sensitive to the details of the simulation.

The highly simplified nature of our model diminishes the need for a detailed study of the seemingly chaotic behavior of neutrino conversions. We postpone such an analysis until more realistic models with multiple dimensions allowing for inhomogeneity have confirmed our findings.
 The right panel of \fref{fig:aniasym} shows that the final number distributions are featureless also in the case of neutrino-antineutrino symmetry breaking. Above $r=10$, the antineutrinos are slightly more equilibrated, which is a feature that arises as the vacuum term dominates the neutrino-neutrino term.

\section{Implications on the effective number of thermally excited neutrino species}
\label{sec:discussion}
In the previous sections, we have seen how an anisotropic setup as well as the introduction of a small asymmetry between neutrinos and antineutrinos can lead to a flavor outcome different than expected.
In order to quantify how large the effect of neutrino flavor conversions can be on the observable quantities, we introduce the effective number of of thermally excited neutrino species:
\begin{equation}
 \label{eq:Neff}
 N_{\rm eff} \equiv \frac{\rho_\nu}{7/8 \rho_\gamma} \left(\frac{11}{7}\right)^{3} \approx \left(\frac{\int dr \; r^3(\rho_{ee} + \bar{\rho}_{ee} + \rho_{xx} + \bar{\rho}_{xx})}{2 \int dr r^3 f_0} + 1 \right) \frac{(11/7)^3}{(T_\gamma/T_{\rm cm})^4}\ .
\end{equation}
Since we do not follow $\nu_y$ in the numerical solution of the QKEs, we assume that $\nu_y$ contributes to $N_{\rm eff}$ as one effective species. We find that the value of $N_{\rm eff}$ in the no conversions case is $3.04596$. Although this value of $N_{\rm eff}$ is quite close to the one obtained by calculations including the full collision term and finite temperature QED effects~\cite{Mangano:2005cc,deSalas:2016ztq,Froustey:2020mcq, Bennett:2020zkv}, we emphasize that this is due to a cancellation of errors. Our estimation neglects finite temperature QED effects, thus decreasing $N_{\rm eff}$; if the collision term had not been approximated and we had included all three neutrinos, we would expect $N_{\rm eff} \approx 3.03$~\cite{Mangano:2005cc,deSalas:2016ztq}. 

Given the approximations involved in our estimation of $N_{\rm eff}$, we focus on 
the difference between the values of $N_{\rm eff}$ after and before flavor conversions, $\Delta N_{\rm eff}$, in the isotropic and anisotropic frameworks. The values of $\Delta N_{\rm eff}$ computed in this way in our simplified framework are listed in \tref{tab:Neff}. 
It is clear that flavor conversions give rise to a slightly larger value of $N_{\rm eff}$. This is to be expected as conversions tend to deplete electron flavor neutrinos, and these are produced more efficiently than the two other flavors of neutrinos. In particular, we find that the correction to $N_{\rm eff}$ due to flavor conversions in NO is comparable to higher order corrections from finite temperature QED effects~\cite{Bennett:2019ewm} and to next to leading order corrections to the weak rates~\cite{Escudero:2020dfa}.

We stress that the variations in $N_{\rm eff}$ reported in \tref{tab:Neff} should be taken with caution given our approximated early universe model with two angle bins only, modelling of flavor mixing in the two flavor approximation and the approximated collision term. However, our findings clearly hint that more work is needed in order to reliably predict the theoretical value of $N_{\rm eff}$ and compare it with observations.

\begin{table}[tbp]
 \centering
 \caption{Change in the effective number of of thermally excited neutrino species ($\Delta N_{\rm eff}$) for the different scenarios analyzed in this work with respect to the case without conversions for isotropic initial conditions (``iso,'' Sec.~\ref{sec:isotropic}) and anisotropic initial conditions (``ani,'' Secs.~\ref{sec:anisotropic} and \ref{sec:symmbreaking}), in NO and IO.}
 \begin{tabular}{l c c c c c}
 \hline\hline
 & NO, iso & NO, ani & NO, ani ($\mu_{\rm ini}=10^{-9}$) & IO, iso/ani\\
 \hline
 $\Delta N_{\rm eff}$ & $0.9\times 10^{-4}$ & $5.0\times 10^{-4}$ & $4.9\times 10^{-4}$ & $5.8\times 10^{-4}$\\
 \hline\hline
 \end{tabular}
 \label{tab:Neff}
\end{table}

\section{Conclusions}
\label{sec:conclusions}

Neutrino self-interactions have been considered to play a negligible role in the flavor evolution of neutrinos in the early universe within standard cosmological scenarios. However, most recent work, carried out in the context of compact astrophysical objects, has shown that the non-linear nature of the neutrino flavor evolution can lead to symmetry breaking effects. This suggests that the eventual presence of small anisotropies in the neutrino field may grow exponentially also in the context of the early universe.

In this work, we carry out the first numerical simulations of the neutrino flavor evolution in an anisotropic setting. Due to the computational limitations intrinsic to the problem, we choose to work within a simplified setup that considers a homogeneous universe with two angle bins: left and right modes for neutrinos and antineutrinos. 
By introducing a small initial asymmetry of $\mathcal{O}(10^{-15})$ between the left and the right modes, we convincingly demonstrate that the flavor evolution is affected in both mass orderings with implications on the effective number of thermally excited neutrino species, $N_{\mathrm{eff}}$. 
In particular, in normal ordering, the flavor evolution in an isotropic scenario is different from the one obtained in the anisotropic case, while the results are comparable in inverted ordering.
These findings are consistent with the predictions of the linear stability analysis applied to our model. 
Interestingly, in normal ordering, we find that the correction to $N_{\rm eff}$ is comparable to higher order corrections from finite temperature QED effects~\cite{Bennett:2019ewm, Bennett:2020zkv}; the value of $N_{\rm eff}$ obtained in this work should be considered with caution given the simplifying assumptions intrinsic to our setup, however it suggests that more work is needed to reliably predict the impact of flavor conversions on cosmological observables.

According to the scenario usually considered for the early universe, the neutrino lepton asymmetry is expected to be of the same order of the baryon one. Since the neutrino self-interaction potential is proportional to the difference between the neutrino and antineutrino number densities, the small neutrino lepton asymmetry has been wrongly used to ignore neutrino self-interactions. However, we find that the lepton asymmetry in the neutrino sector can be amplified with time in the presence of anisotropies in the neutrino field, due to the non-linear nature of the neutrino flavor evolution in the presence of neutrino self-interactions. However, we stress that the amplification of the neutrino-antineutrino asymmetry also occurs in the isotropic case and it is an effect independent from the breaking of isotropy.

Our findings can have very interesting implications for non-standard scenarios of the early universe, such as low temperature reheating, mixing between active and extra sterile neutrino states, or flavor mixing in the presence of a large lepton asymmetry~\cite{Kawasaki:2000en, Ichikawa:2005vw, deSalas:2015glj, Hasegawa:2019jsa, Kainulainen:1990ds, Hannestad:2015tea, Dolgov:2002ab, Wong:2002fa, Abazajian:2002qx, Hannestad:2012ky}.
In particular, low temperature reheating is in many ways the perfect place to expect anisotropic conversions~\cite{Kawasaki:2000en, Ichikawa:2005vw, deSalas:2015glj, Hasegawa:2019jsa}. Assuming that neutrinos are not directly produced in the decay giving rise to reheating, they may be produced via weak Standard Model processes leading to an overabundance of $\nu_e$ and $\bar\nu_e$, as long as neutrinos are not in full equilibrium. This could trigger flavor instabilities and a large difference between the neutrino flavors could lead to the growth of significant anisotropies. Furthermore, if the reheating temperature is low enough, the spectral distortions and anisotropies developed in the neutrino sector could survive the weak decoupling epoch and leave an imprint at later times.
For active-sterile mixing, the asymmetries introduced between the different species of active neutrinos are quite small, if flavor conversions occur before weak decoupling. However, if conversions take place close to $T_\gamma\simeq 1$~MeV, there could be a significant effect due to anisotropic conversions that could even leak into the sterile sector.
Finally, the amplification of the neutrino-antineutrino asymmetry due to flavor conversions raises the question of whether active-active neutrino conversions in the early universe are not only anisotropic and inhomogeneous, but also chaotic, see e.g.~Ref.~\cite{Keister:2014ufa}.

In conclusion, neutrino flavor mixing may possibly play an even more profound role in the early universe than what currently assumed. Our results show without a doubt that it is imperative to revise the neutrino flavor conversion physics in the light of our more advanced understanding of neutrino propagation in dense media to gauge possible effects on the cosmological observables.

\acknowledgments
We are grateful to Steen Hannestad for helpful discussions and to the anonymous referee for insightful comments and suggestions. This project has received funding from the the Villum Foundation (Project No.~13164), the Danmarks Frie Forskningsfonds (Project No.~8049-00038B), the European Union's Horizon 2020 research and innovation program under the Marie Sklodowska-Curie grant agreement No.~847523 (``INTERACTIONS''), the Knud H\o jgaard Foundation, and the Deutsche Forschungsgemeinschaft through Sonderforschungbereich
SFB~1258 ``Neutrinos and Dark Matter in Astro- and Particle Physics'' (NDM).

\appendix

\section{Collision term}
\label{app:collterm}

 The collision term has the following general form:
 \begin{equation}
 \label{eq:collgen}
 \mathcal{C} = \mathcal{C}_{\rm gain} - \mathcal{C}_{\rm loss}\ ,
 \end{equation}
 with a gain term and a loss term. In general, for a process
 \begin{equation}
 \label{eq:process}
 a(\mathbf{p}) + b(\mathbf{k}) \leftrightarrow c(\mathbf{p}^{\prime}) + d(\mathbf{k}^{\prime})\ ,
 \end{equation}
 we wish to know the number of particles of species $a$ with momentum $\mathbf{p}$ that are lost and gained due to that process. The gained number of particles of species $a$ is proportional to $f_c(\mathbf{p}^\prime) f_d(\mathbf{k}^\prime)$ summed over all $\mathbf{p}^\prime$ and $\mathbf{k}^\prime$, while the lost number of particles of species $a$ is proportional to $f_a(\mathbf{p}) f_b(\mathbf{k})$ summed over all $\mathbf{k}$. Hence we have:
 \begin{equation}
 \label{eq:gainterm}
 \mathcal{C}_{\rm gain} \propto \sum_{\mathbf{p}^{\prime},\mathbf{k}^{\prime}} f_c(\mathbf{p}^{\prime}) f_d(\mathbf{k}^{\prime})
 \end{equation}
 \begin{equation}
 \label{eq:lossterm}
 \mathcal{C}_{\rm loss} \propto \sum_{\mathbf{k}} f_a(\mathbf{p}) f_b(\mathbf{k})\ ,
 \end{equation}
 when Pauli blocking is neglected. For the neutrino density matrices, the simple products are replaced by more complicated structures and anticommutators which we have included for each of the processes in the following.

In order to find an approximation for the collision term that retains as much physics as possible, while keeping it light enough that it can be evaluated without a high numerical cost, we adopt an approach similar to the A/S approximation of Ref.~\cite{Hannestad:2015tea} with the important difference that we are now dealing with active-active conversions and not active-sterile conversions.
In order to keep an appropriate dependence on the distribution functions, we use the expressions derived in Ref.~\cite{Blaschke:2016xxt}. We neglect Pauli blocking and integrate out the collision rate for each term. In the following, we consider the various contributions to the collision term, and then validate our approximation.

\subsection{Neutrino-electron scattering}

For neutrino scattering with electrons, there are three different contributions with different dependence on the distribution functions.
The structure in terms of distributions for the first part is
\begin{equation}
 \mathcal{C} \propto \left\{Y_L f_3 Y_L, 1 \right\} - \left\{Y_L Y_L, f \right\}\ ,
\end{equation}
where $f$ is the distribution of the incoming neutrino, $f_3$ is the distribution of the outgoing neutrino and 
\begin{equation}
 Y_L =
 \begin{pmatrix}
 \frac{1}{2} + \sin^2\theta_W & 0 \\ 0 & - \frac{1}{2} + \sin^2\theta_W 
 \end{pmatrix}\ .
\end{equation}
Integrating over the matrix elements, we write the approximate collision term as
\begin{align}
 \mathcal{C}_{ee} &= \Gamma_1 \left(\frac{1}{2} + \sin^2\theta_W\right)^2 (f(T_\gamma,\pi_e)-\rho_{ee})\ , \\
 \mathcal{C}_{xx} &= \Gamma_1 \left(-\frac{1}{2} + \sin^2\theta_W\right)^2 (f(T_\gamma,\pi_x) - \rho_{xx})\ , \\
 \mathcal{C}_{ex} &= \Gamma_1 \left(-\frac{1}{4}+\sin^4\theta_W\right) u_{ex} f_0 - \left(\frac{1}{4} + \sin^4\theta_W\right) \rho_{ex}\ ,
\end{align}
where
\begin{equation}
 \label{eq:Gamma1}
 \Gamma_1 = C_1 G_{
\rm F}^2 p T_{\rm cm}^4\ ,
\end{equation}
\begin{equation}
 u_{\alpha\beta} = \frac{\int dp \;p \;\rho_{\alpha\beta}}{\int dp \; p \;f_0}\ ,
\end{equation}
the equilibrium distribution for $\nu_\alpha$ is given by
\begin{equation*}
 f(T_{\rm eq},\pi_\alpha) = \frac{1}{\exp(p/T_{\rm eq} - \pi_\alpha/T_{\rm eq}) + 1}\ ,
\end{equation*}
and $f_0 = f(T_{\rm cm},0)$ is the Fermi-Dirac distribution with zero chemical potential. Notice that $\pi_\alpha$ and $\bar{\pi}_\alpha$ denotes the pseudo-chemical potentials for neutrinos and antineutrinos which conserves the number density of each neutrino and antineutrino.
 This is done by determining $\pi_\alpha$ such that $\int dp \;p^3 (f(T_\gamma,\pi_\alpha) - \rho_{\alpha\alpha}) = 0$ and similarly for $\bar{\pi}_\alpha$.
The chemical potential $\mu_\alpha$ satisfies the condition $\bar{\mu}_\alpha = -\mu_\alpha$ and only conserves the lepton number. 

For the second term, the distribution function dependence is:
\begin{equation}
 \mathcal{C} \propto \left\{Y_R f_3 Y_R, 1 \right\} - \left\{Y_R Y_R, f \right\}\ ,
\end{equation}
where 
\begin{equation}
 Y_R = \sin^2 \theta_W I\ .
\end{equation}
This gives the approximation
\begin{align}
 \mathcal{C}_{ee} &= \Gamma_2 \sin^4\theta_W (f(T_\gamma,\pi_e)-\rho_{ee})\ , \\
 \mathcal{C}_{xx} &= \Gamma_2 \sin^4\theta_W (f(T_\gamma,\pi_x) - \rho_{xx})\ ,\\
 \mathcal{C}_{ex} &= \Gamma_2 \sin^4\theta_W (u_{ex} f_0 - \rho_{ex})\ , 
\end{align}
where
\begin{equation}
 \label{eq:Gamma2}
 \Gamma_2 = C_2 G_{
\rm F}^2 p T_{\rm cm}^4 
\end{equation}
as for the first term.

Finally, the third term has the form
\begin{equation}
 \mathcal{C} \propto \left\{ Y_L f_3 Y_R, 1 \right\} + \left\{ Y_R f_3 Y_L, 1 \right\} - \left\{ Y_L Y_R, f \right\} - \left\{ Y_R Y_L, f \right\}\ .
\end{equation}
Again, we make similar approximations:
\begin{align}
 \mathcal{C}_{ee} &= \Gamma_3 \left(\frac{1}{2} + s^2_{\theta_W}\right) s^2_{\theta_W} (f(T_\gamma,\pi_e)-\rho_{ee})\ , \\
 \mathcal{C}_{xx} &= \Gamma_3 \left(-\frac{1}{2} + s^2_{\theta_W}\right) s^2_{\theta_W} \left(f(T_\gamma,\pi_x) - \rho_{xx}\right)\ ,\\
 \mathcal{C}_{ex} &= \Gamma_3 s^4_{\theta_W}\left( u_{ex} f_0 - \rho_{ex}\right)\ , 
\end{align}
where
\begin{equation}
 \label{eq:Gamma3}
 \Gamma_3 = C_3 G_{
\rm F}^2 p T_{\rm cm}^4\ ,
\end{equation}
and $s_{\theta_W} = \sin(\theta_W)$.

Combining all three terms gives
\begin{align}
 \mathcal{C}_{ee} &= \left( \Gamma_1 \left(\tfrac{1}{2} + s^2_{\theta_W}\right)^2 + \Gamma_2 s^4_{\theta_W} + \Gamma_3 \left(\tfrac{1}{2} + s^2_{\theta_W}\right)s^2_{\theta_W} \right) (f(T_\gamma,\pi_e) - \rho_{ee})\ ,\\
 \mathcal{C}_{xx} &= \left( \Gamma_1 \left(-\tfrac{1}{2} + s^2_{\theta_W}\right)^2 + \Gamma_2 s^4_{\theta_W} + \Gamma_3 \left(-\tfrac{1}{2} + s^2_{\theta_W}\right) s^2_{\theta_W} \right) (f(T_\gamma,\pi_x) - \rho_{xx})\ , \\
 \mathcal{C}_{ex} &= \left(\Gamma_1 \left(-\tfrac{1}{4} + s^4_{\theta_W}\right) + \Gamma_2 s^4_{\theta_W} + \Gamma_3 s^4_{\theta_W} \right) u_{ex} f_0 - \left(\Gamma_1 \left(\tfrac{1}{4} + s^4_{\theta_W}\right) + \Gamma_2 s^4_{\theta_W} + \Gamma_3 s^4_{\theta_W}\right) \rho_{ex}\ .
\end{align}
Since these are scatterings with electrons, the conserved quantities are the lepton number for each lepton flavor as well as the number density of each neutrino flavor. 
The energy density of neutrinos is not conserved as scatterings with electrons will push the neutrino energy distribution towards the photon temperature. For this reason, $T_\gamma$ is used in the equilibrium distributions. The pseudo-chemical potentials are determined for $\nu_e$, $\bar\nu_e$, $\nu_x$, and $\bar\nu_x$ in order to conserve the number densities of all neutrinos independently. The conservation of all number densities automatically leads to the lepton number conservation.

 The temperature used for calculating $\Gamma_i$ has been chosen to be $T_{\rm cm}$. This choice reflects the neutrino temperature up to a small correction, but it is an underestimation for electrons and positrons at low temperatures. However, this approximation reproduces the results from the full collision term quite well for scatterings as the test at the end of this appendix shows.

The coefficients for $\Gamma_i$ are shown in \tref{tab:Ce} for some representative values of $m_e/T_{\rm cm}$~\footnote{A modified version of the code for calculating repopulation and damping coefficients of Ref.~\cite{Hannestad:2015tea} was used for this calculation.}.
The asymptotic value of $m_e/T_{\rm cm}$ is a good approximation above 0.5~MeV, and the value of $\Gamma_i$ is suppressed at lower temperatures by the $T_{\rm cm}^5$ dependence such that $\Gamma_i/H$ decreases as $T_{\rm cm}^3$.
Hence we take the values at $m_e/T_{\rm cm} \approx 0$ for all temperatures.

\begin{table}[htbp]
 \centering
 \caption{Coefficients for neutrino-electron scattering.}
 \begin{tabular}{c c c c c c}
 \hline\hline
 $m_e/T_{\rm cm}$ & $C_1$ & $C_2$ & $C_3$ & $T_{\rm cm}$ & $T_{\rm cm}^3$\\
 \hline
 0.0 & 0.489 & 0.163 & 0 & $\infty$ & $\infty$\\
 0.5 & 0.471 & 0.159 & -0.004 & 1.022~MeV & 1.067~MeV$^3$\\
 1.0 & 0.425 & 0.147 & -0.014 & 0.511~MeV & 0.133~MeV$^3$\\
 2.0 & 0.294 & 0.109 & -0.028 & 0.256~MeV & 0.017~MeV$^3$\\
 4.0 & 0.093 & 0.040 & -0.021 & 0.128~MeV & 0.002~MeV$^3$\\
 \hline\hline
 \end{tabular}
 \label{tab:Ce}
\end{table}

\subsection{Neutrino-positron scattering}

Compared to neutrino-electron scattering, the neutrino-positron scattering terms have the same structure, but the constants are swapped such that $C_1 \leftrightarrow C_2$. Hence, we obtain
\begin{align}
 \mathcal{C}_{ee} &= \left( \Gamma_2 \left(\tfrac{1}{2} + s^2_{\theta_W}\right)^2 + \Gamma_1 s^4_{\theta_W} + \Gamma_3 \left(\tfrac{1}{2} + s^2_{\theta_W}\right)s^2_{\theta_W} \right) (f(T_\gamma,\pi_e) - \rho_{ee})\ ,\\
 \mathcal{C}_{xx} &= \left( \Gamma_2 \left(-\tfrac{1}{2} + s^2_{\theta_W}\right)^2 + \Gamma_1 s^4_{\theta_W} + \Gamma_3 \left(-\tfrac{1}{2} + s^2_{\theta_W}\right) s^2_{\theta_W} \right) (f(T_\gamma,\pi_x) - \rho_{xx})\ , \\
 \mathcal{C}_{ex} &= \left(\Gamma_2 \left(-\tfrac{1}{4} + s^4_{\theta_W}\right) + \Gamma_1 s^4_{\theta_W} + \Gamma_3 s^4_{\theta_W} \right) u_{ex} f_0 - \left(\Gamma_2 \left(\tfrac{1}{4} + s^4_{\theta_W}\right) + \Gamma_1 s^4_{\theta_W} + \Gamma_3 s^4_{\theta_W}\right) \rho_{ex}\ ,
\end{align}
where the coefficients are reported in \tref{tab:Ce}.

\subsection{Electron-positron annihilation}

For annihilation processes, the antineutrino distribution function is present in the loss term rather than in the gain term. Hence,
\begin{equation}
 \mathcal{C} \propto \left\{Y_L Y_L, 1\right\} - \{Y_L \bar{f}_3 Y_L,f\}\ ,
\end{equation}
where $\bar{f}_3$ is the distribution function of antineutrinos.
We approximate it as
\begin{align}
 \mathcal{C}_{ee} &= \Gamma_{1a} \left(\tfrac{1}{2} + s^2_{\theta_W} \right)^2 \left(\left(\tfrac{T_\gamma}{T_{\rm cm}}\right)^4 f(T_\gamma,0) - \bar{u}_{ee}\rho_{ee}\right) - \Gamma_{1a}(-\tfrac{1}{4}+s^4_{\theta_W}){\rm Re}(\bar{u}_{ex} \rho_{ex}^*)\ ,\\ 
 \mathcal{C}_{xx} &= \Gamma_{1a} \left(-\tfrac{1}{2} + s^2_{\theta_W}\right)^2 \left(\left(\tfrac{T_\gamma}{T_{\rm cm}}\right)^4 f(T_\gamma,0) - \bar{u}_{xx} \rho_{xx}\right) - \Gamma_{1a} (-\tfrac{1}{4}+s^4_{\theta_W}){\rm Re}(\bar{u}_{ex}^* \rho_{ex})\ ,\\
 \mathcal{C}_{ex} &= - \tfrac{1}{2} \Gamma_{1a} (-\tfrac{1}{4} + s^4_{\theta_W}) \bar{u}_{ex} (\rho_{ee} + \rho_{xx}) - \Gamma_{1a}(\tfrac{1}{4} + s^4_{\theta_W}) \rho_{ex}\ ,
\end{align}
where
\begin{equation}
 \bar{u}_{\alpha\beta} = \frac{\int dp \; p \;\bar{\rho}_{\alpha\beta}}{\int dp \;p \; f_0}\ ,
\end{equation}
and the factor $\left(\frac{T_\gamma}{T_{\rm cm}}\right)^4$ accounts for the higher temperature of the electron-positron-photon plasma.
In $\mathcal{C}_{ex}$, we put $\bar{u}_{xx}=1$.

When solving the QKEs with this collision term and in the presence of a lepton asymmetry, the neutrino spectra are not pushed towards the equilibrium distributions $f(T_\gamma,\mu)$ at high temperatures as they should. Instead they are pushed towards off-set distributions with zero chemical potential. To resolve this problem, we replace
\begin{equation*}
 \left(\frac{T_\gamma}{T_{\rm cm}}\right)^4 f(T_\gamma,0) - \bar{u}_{xx} \rho_{xx} \qquad {\rm by} \qquad
 \left(\frac{T_\gamma}{T_{\rm cm}}\right)^4 f(T_\gamma,\mu_x) - \rho_{xx}\ ,
\end{equation*}
where the chemical potential $\mu_x$ is determined such that the lepton asymmetry is conserved for each neutrino flavor. Thus, the equations we use are
\begin{align}
 \mathcal{C}_{ee} &= \Gamma_{1a} \left(\tfrac{1}{2} + s^2_{\theta_W} \right)^2 \left(\left(\tfrac{T_\gamma}{T_{\rm cm}}\right)^4 f(T_\gamma,\mu_e) - \rho_{ee}\right) - \Gamma_{1a}(-\tfrac{1}{4}+s^4_{\theta_W}){\rm Re}(\bar{u}_{ex} \rho_{ex}^*)\ ,\\ 
 \mathcal{C}_{xx} &= \Gamma_{1a} \left(-\tfrac{1}{2} + s^2_{\theta_W}\right)^2 \left(\left(\tfrac{T_\gamma}{T_{\rm cm}}\right)^4 f(T_\gamma,\mu_x) - \rho_{xx}\right) - \Gamma_{1a} (-\tfrac{1}{4}+s^4_{\theta_W}){\rm Re}(\bar{u}_{ex}^* \rho_{ex})\ ,\\
 \mathcal{C}_{ex} &= - \tfrac{1}{2} \Gamma_{1a} (-\tfrac{1}{4} + s^4_{\theta_W}) \bar{u}_{ex} (\rho_{ee} + \rho_{xx}) - \Gamma_{1a}(\tfrac{1}{4} + s^4_{\theta_W}) \rho_{ex}\ .
\end{align}
On the diagonal terms, we find a modified relaxation time approximation and, in addition, a term which is due to mixed neutrinos and mixed antineutrinos annihilating into electron/positron pairs. On the off-diagonal, there is another term which is again due to mixed antineutrino annihilating with neutrinos.

For the second term, the distribution function dependence is:
\begin{equation}
 \mathcal{C} \propto \left\{Y_R Y_R, 1 \right\} - \left\{Y_R \bar{f}_3 Y_R, f \right\}\ .
\end{equation}
With the same modification as for the first term, the approximation is
\begin{align}
 \mathcal{C}_{ee} &= \Gamma_{2a}\sin^4\theta_W \left(\left(\tfrac{T_\gamma}{T_{\rm cm}}\right)^4 f(T_\gamma,\mu_e)- \rho_{ee} - {\rm Re}(\bar{u}_{ex} \rho_{ex}^*)\right)\ ,\\
 \mathcal{C}_{xx} &= \Gamma_{2a}\sin^4\theta_W \left(\left(\tfrac{T_\gamma}{T_{\rm cm}}\right)^4 f(T_\gamma,\mu_x) - \rho_{xx}- {\rm Re}(\bar{u}_{ex}^* \rho_{ex})\right)\ ,\\ 
 \mathcal{C}_{ex} &= -\Gamma_{2a}\sin^4\theta_W \left( \frac{1}{2} \bar{u}_{ex} (\rho_{ee} + \rho_{xx}) + \rho_{ex} \right)\ . 
\end{align}
Again we find similar corrections due to coherence in the neutrino and antineutrino fields.

Finally, the third term has the form
\begin{align}
 \mathcal{C} &= \left\{ Y_L Y_R, 1\right\} + \left\{ Y_R Y_L, 1 \right\} - \left\{ Y_L \bar{f}_3 Y_R, f \right\} - \left\{ Y_R \bar{f}_3 Y_L, f \right\}\ ,
\end{align}
which gives the modified approximation
\begin{align}
 \mathcal{C}_{ee} &= \Gamma_{3a} (\tfrac{1}{2} + s^2_{\theta_W}) s^2_{\theta_W} \left( \left(\tfrac{T_\gamma}{T_{\rm cm}}\right)^4 f(T_\gamma,\mu_e) - \rho_{ee}\right) - \Gamma_{3a} s^4_{\theta_W} {\rm Re}(\bar{u}_{ex} \rho_{ex}^*)\ ,\\
\mathcal{C}_{xx} &= \Gamma_{3a} (-\tfrac{1}{2} + s^2_{\theta_W}) s^2_{\theta_W} \left( \left(\tfrac{T_\gamma}{T_{\rm cm}}\right)^4 f(T_\gamma,\mu_x) - \rho_{xx}\right) - \Gamma_{3a} s^4_{\theta_W} {\rm Re}(\bar{u}_{ex}^* \rho_{ex})\ ,\\
 \mathcal{C}_{ex} &= -\tfrac{1}{2} \Gamma_{3a} s^4_{\theta_W} \bar{u}_{ex} (\rho_{ee} + \rho_{xx}) - \Gamma_{3a} s^4_{\theta_W} \rho_{ex}\ .
\end{align}
Summing over all three contributions, we find
\begin{align}
 \mathcal{C}_{ee} &= \left( \Gamma_{1a} \left(\tfrac{1}{2} + s^2_{\theta_W}\right)^2 + \Gamma_{2a} s^4_{\theta_W} + \Gamma_{3a} \left(\tfrac{1}{2} + s^2_{\theta_W}\right)s^2_{\theta_W} \right) \left( \left(\tfrac{T_\gamma}{T_{\rm cm}}\right)^4 f(T_\gamma,\mu_e) - \rho_{ee}\right) \nonumber\\
& \qquad\qquad - \left(\Gamma_{1a} \left(-\tfrac{1}{4} + s^4_{\theta_W}\right) + \Gamma_{2a} s^4_{\theta_W} + \Gamma_{3a} s^4_{\theta_W}\right) {\rm Re}\left(\bar{u}_{ex} \rho_{ex}^* \right)\ , \label{eq:Cee_a}\\
 \mathcal{C}_{xx} &= \left( \Gamma_{1a} \left(-\tfrac{1}{2} + s^2_{\theta_W}\right)^2 + \Gamma_{2a} s^4_{\theta_W} + \Gamma_{3a} \left(-\tfrac{1}{2} + s^2_{\theta_W}\right) s^2_{\theta_W} \right) \left( \left(\tfrac{T_\gamma}{T_{\rm cm}}\right)^4 f(T_\gamma,\mu_x) - \rho_{xx}\right)\nonumber\\
& \qquad\qquad - \left(\Gamma_{1a} \left(-\tfrac{1}{4} + s^4_{\theta_W}\right) + \Gamma_{2a} s^4_{\theta_W} + \Gamma_{3a} s^4_{\theta_W}\right) {\rm Re}\left(\bar{u}_{ex} \rho_{ex}^* \right)\ ,\label{eq:Cmm_a}\\
 \mathcal{C}_{ex} &= - \left(\Gamma_{1a} \left(\tfrac{1}{4} + s^4_{\theta_W}\right) + \Gamma_{2a} s^4_{\theta_W} + \Gamma_{3a} s^4_{\theta_W}\right) \rho_{ex} \nonumber\\
& \qquad \qquad -\frac{1}{2} \left(\Gamma_{1a} \left(-\tfrac{1}{4} + s^4_{\theta_W}\right) + \Gamma_{2a} s^4_{\theta_W} + \Gamma_{3a} s^4_{\theta_W} \right) \bar{u}_{ex} (\rho_{ee} + \rho_{xx})\ . \label{eq:Cem_a}
\end{align}
The first line for each element of the collision term is a relaxation time approximation, and the second line for each element contains contributions from coherent neutrinos and antineutrinos which are often neglected in the literature. The coefficients for $\Gamma_{ia}$ are shown in \tref{tab:Cea} for some representative values of $m_e/T_{\rm cm}$, and as for the scattering terms, we use the approximation $m_e/T_{\rm cm}=0$ in our calculations.
\begin{table}[htbp]
 \centering
 \caption{Coefficients for neutrino-electron scattering.}
 \begin{tabular}{c c c c c c}
 \hline\hline
 $m_e/T_{\rm cm}$ & $C_{1a}$ & $C_{2a}$ & $C_{3a}$ & $T_{\rm cm}$ & $T_{\rm cm}^3$\\
 \hline
 0.0 & 0.163 & 0.220 & 0 & $\infty$ & $\infty$\\
 0.5 & 0.158 & 0.217 & -0.009 & 1.022~MeV & 1.067~MeV$^3$\\
 1.0 & 0.145 & 0.207 & -0.029 & 0.511~MeV & 0.133~MeV$^3$\\
 2.0 & 0.097 & 0.160 & -0.056 & 0.256~MeV & 0.017~MeV$^3$\\
 4.0 & 0.019 & 0.041 & -0.021 & 0.128~MeV & 0.002~MeV$^3$\\
 \hline\hline
 \end{tabular}
 \label{tab:Cea}
\end{table}

\subsection{All electron/positron contributions to the collision term}

Collecting all the terms from scattering and annihilation, and defining a number of constants, we find
\begin{align}
 \mathcal{C}_{ee} &= \Gamma_{s,e} (f(T_\gamma,\pi_{e}) - \rho_{ee}) + \Gamma_{a,e} \left( \left(\tfrac{T_\gamma}{T_{\rm cm}}\right)^4 f(T_\gamma,\mu_{e}) - \rho_{ee} \right) - \Gamma_G {\rm Re}\left( \bar{u}_{ex} \rho_{ex}^* \right)\ ,\\
 \mathcal{C}_{xx} &= \Gamma_{s,x} (f(T_\gamma,\pi_{x}) - \rho_{xx}) + \Gamma_{a,x} \left( \left(\tfrac{T_\gamma}{T_{\rm cm}}\right)^4 f(T_\gamma,\mu_{x}) - \rho_{xx}\right) - \Gamma_G {\rm Re}\left( \bar{u}_{ex} \rho_{ex}^* \right)\ ,\\
 \mathcal{C}_{ex} &= - D \rho_{ex} + d\; u_{ex} f_0 - C \bar{u}_{ex} (\rho_{ee} + \rho_{xx})\ ,
\end{align}
where
\begin{align}
 \Gamma_{s,\alpha} &= \left(\Gamma_{1} +\Gamma_{2}\right) \left( \left(\pm\tfrac{1}{2} + s^2_{\theta_W}\right)^2 + s^4_{\theta_W} \right) + 2 \Gamma_{3} \left(\pm\tfrac{1}{2} + s^2_{\theta_W}\right)s^2_{\theta_W}\ ,\\
\Gamma_{a,\alpha} &= \Gamma_{1a} \left(\pm\tfrac{1}{2} + s^2_{\theta_W}\right)^2 + \Gamma_{2a} s^4_{\theta_W} + \Gamma_{3a} \left(\pm\tfrac{1}{2} + s^2_{\theta_W}\right)s^2_{\theta_W}\ ,\\
\Gamma_G &= \Gamma_{1a} \left(-\tfrac{1}{4} + s^4_{\theta_W}\right) + \left( \Gamma_{2a} + \Gamma_{3a} \right) s^4_{\theta_W}\ ,\\
D &= \left(\Gamma_{1} + \Gamma_{2} \right) \left(\tfrac{1}{4} + 2 s^4_{\theta_W}\right) + \left( 2 \Gamma_{3} + \Gamma_{2a} + \Gamma_{3a} \right) s^4_{\theta_W} + \Gamma_{1a} \left(\tfrac{1}{4} + s^4_{\theta_W}\right)\ ,\\
d &= \left( \Gamma_{1} + \Gamma_{2} \right) \left(-\tfrac{1}{4} + 2 s^4_{\theta_W} \right) + 2 \Gamma_{3} s^4_{\theta_W}\ , \\
C &= \frac{1}{2} \left(\Gamma_{1a} \left(-\tfrac{1}{4} + s^4_{\theta_W}\right) + \Gamma_{2a} s^4_{\theta_W} + \Gamma_{3a} s^4_{\theta_W} \right)\ ,
\end{align}
and plus signs are for $\alpha=e$ while minus signs are for $\alpha = x$.
These can also be expressed as
\begin{align}
 \Gamma_{s,\alpha} &= y_{s,\alpha} G_{
\rm F}^2 p T_{\rm cm}^4,&\qquad \Gamma_{a,\alpha} &= y_{a,\alpha} G_{
\rm F}^2 p T_{\rm cm}^4\ ,\label{eq:Gammaab}\\
\Gamma_G &= y_{G} G_{
\rm F}^2 p T_{\rm cm}^4,&\qquad D &= y_D G_{
\rm F}^2 p T_{\rm cm}^4\ ,\\
d &= y_{d} G_{
\rm F}^2 p T_{\rm cm}^4, &\qquad C &= y_{C} G_{
\rm F}^2 p T_{\rm cm}^4\ ,\label{eq:dC}
\end{align}
where $y_D = 0.296$ and the other values for $y$ are found in the last row of \tref{tab:colltest}.

The equilibrium distribution functions $f(T_\gamma,\pi_{\alpha})$ relevant for scattering processes are defined such that the number integral over $\rho_{\alpha\alpha}$ is conserved. In principle, this implies the conservation of the lepton number as well. However, if the individual number densities are conserved to an absolute precision $\Delta n$ for neutrinos and $\Delta \bar{n}$ for antineutrinos, then the relative error on the lepton asymmetry $\Delta L/L \lesssim (\Delta n + \Delta \bar{n})/(n-\bar{n})$ could be order one or larger.
This can be accounted for by determining the two pseudo-chemical potentials $\pi_\nu$ and $\bar{\pi}_{\nu}$ simultaneously conserving the number integrals over $\rho_{\alpha\alpha} \pm \bar{\rho}_{\alpha \alpha}$. The initial guesses for $\pi_{\nu}$ and $\bar{\pi}_{\nu}$ are determined by conservation of the individual number densities, and afterwards they are refined using a Newton-Raphson method.

\subsection{Neutrino-neutrino scattering}
Neutrino-neutrino interactions are more challenging to treat as they do not involve any particles that are assumed to be in equilibrium; however, following the same approach as above is still straightforward. The distribution function dependence for neutrino-neutrino scattering without Pauli blocking factors is
\begin{equation}
 \mathcal{C} \propto \left\{ \left(\Tr{f_1 } + f_1 \right)f_3,1\right\} -\left\{ \Tr{ f_2} + f_2, f\right\}\ .
\end{equation}
These two terms can be approximated by
\begin{align}
 \mathcal{C} &= \Gamma_{\nu\nu}
 \begin{pmatrix}
 (2u_{ee} + u_{xx}) u_{ee} + u_{ex} u_{ex}^* & 2(u_{ee} + u_{xx}) u_{ex}\\
 2(u_{ee} + u_{xx} ) u_{ex}^* & (u_{ee} + 2 u_{xx} ) u_{xx} + u_{ex}^* u_{ex}
 \end{pmatrix} f_0\\
 &\quad - \Gamma_{\nu\nu}
 \begin{pmatrix}
 (2u_{ee}+u_{xx} ) \rho_{ee} + {\rm Re}(u_{ex} \rho_{ex}^*) & \frac{3}{2}(u_{ee}+ u_{xx} ) \rho_{ex} + \frac{1}{2}u_{ex} (\rho_{ee} + \rho_{xx}) \\
 \frac{3}{2}(u_{ee}+ u_{xx} ) \rho_{ex}^* + \frac{1}{2}u_{ex}^* (\rho_{ee} + \rho_{xx}) & (u_{ee}+2u_{xx} ) \rho_{xx} + {\rm Re}(u_{ex} \rho_{ex}^*) \nonumber
 \end{pmatrix}\ .
\end{align}
The diagonal elements integrate to zero, so the number density of each flavor is conserved.

As for the electron-positron annihilation term, this approximation has the problem that an equilibrium distribution is not reproduced at high temperatures. For this reason, we instead use the approximation

\begin{align}
 \mathcal{C} &= \Gamma_{\nu\nu}
 \begin{pmatrix}
 (2u_{ee} + u_{xx}) f_{\nu_e} + {\rm Re}(u_{ex}^* f_{ex}) & \frac{3}{2}(u_{ee} + u_{xx} ) f_{ex} + \frac{1}{2} u_{ex}(f_{\nu_e}+f_{\nu_x})\\
 \frac{3}{2}(u_{ee} + u_{xx} ) f_{ex}^* + \frac{1}{2} u_{ex}^*(f_{\nu_e}+f_{\nu_x}) & (u_{ee} + 2 u_{xx} ) f_{\nu_x} + {\rm Re}(u_{ex}^* f_{ex}) \end{pmatrix}\nonumber\\
 &\quad - \Gamma_{\nu\nu}
 \begin{pmatrix}
 (2u_{ee}+u_{xx} ) \rho_{ee} + {\rm Re}(u_{ex}^* \rho_{ex}) & \frac{3}{2}(u_{ee}+ u_{xx} ) \rho_{ex} + \frac{1}{2}u_{ex} (\rho_{ee} + \rho_{xx}) \\
 \frac{3}{2}(u_{ee}+ u_{xx} ) \rho_{ex}^* + \frac{1}{2}u_{ex}^* (\rho_{ee} + \rho_{xx}) & (u_{ee}+2u_{xx} ) \rho_{xx} + {\rm Re}(u_{ex}^* \rho_{ex}) 
 \end{pmatrix}\ ,
\end{align}
where $f_{\nu_\alpha} = f(T_{\nu_\alpha}, \pi_{\nu_\alpha})$ with the pseudo-chemical potentials $\pi_{\nu_\alpha}$ and $\bar{\pi}_{\nu_\alpha}$ and temperatures $T_{\nu_\alpha}$ and $T_{\bar{\nu}_\alpha}$ which are determined to conserve the number and energy density of every flavor of neutrinos and antineutrinos. 
 This is done by ensuring that $\int dp \;p^3 (f_{\nu_\alpha} - \rho_{\alpha\alpha}) = 0$ and $\int dp \;p^4 (f_{\nu_\alpha} - \rho_{\alpha\alpha}) = 0$.
For the off-diagonal term, 
\begin{equation*}
 f_{ex} = (a_x+b_x p) f_0 + i (a_y + b_y p) f_0\ .
\end{equation*}
Here $a_x$ and $b_x$ are determined such that $\int dp^3 p (a_x+b_x p) f_0 = \int dp^3 p {\rm Re}(\rho_{ex})$ and $\int dp^3 p^2 (a_x+b_x p) f_0 = \int dp^3 p^2 {\rm Re}(\rho_{ex})$.

\subsection{Antineutrino-neutrino collisions}

The form of the antineutrino-neutrino annihilation and scattering term is
\begin{equation}
 \mathcal{C} \propto \left\{ (\Tr{\bar{f}_1} + \bar{f}_1)f_3 + (\Tr{f_3\bar{f}_1} + f_3\bar{f}_1), 1\right\} - \left\{ \Tr{\bar{f}_2} + 4 \bar{f}_2, f\right\}\ ,
\end{equation}
and our approximation to the collision term is
\begin{align}
 \mathcal{C}_{ee} &= \Gamma_{\nu\bar{\nu}} (4 u_{ee} \bar{u}_{ee} + (u_{ee} + u_{xx}) \bar{u}_{xx} + 4 {\rm Re}(u_{ex} \bar{u}_{ex}^*) ) f_0 \nonumber\\
&\quad- \Gamma_{\nu\bar{\nu}} ( (5\bar{u}_{ee} + \bar{u}_{xx} ) \rho_{ee} + 4 {\rm Re} (\bar{u}_{ex} \rho_{ex}^*))\ ,\\
\mathcal{C}_{xx} &= \Gamma_{\nu\bar{\nu}} (4 u_{xx} \bar{u}_{xx} + (u_{ee} + u_{xx}) \bar{u}_{ee} + 4 {\rm Re}( u_{ex}^* \bar{u}_{ex})) f_0\nonumber\\
&\quad - \Gamma_{\nu\bar{\nu}} ((5\bar{u}_{xx} + \bar{u}_{ee} ) \rho_{xx} + 4{\rm Re}(\bar{u}_{ex}^* \rho_{ex}))\ ,\\
\mathcal{C}_{ex} &= \Gamma_{\nu\bar{\nu}}( 2u_{ex} (\bar{u}_{ee} + \bar{u}_{xx} ) + \bar{u}_{ex} (u_{ee} + u_{xx})) f_0\nonumber\\
&\quad - \Gamma_{\nu\bar{\nu}} ( 3(\bar{u}_{ee} + \bar{u}_{xx} ) \rho_{ex} + 2 \bar{u}_{ex} (\rho_{ee} + \rho_{xx}))\ .
\end{align}
The diagonal terms only integrate to zero when taking the difference between neutrinos and antineutrinos, so the number densities are not conserved, but the lepton asymmetries are. Furthermore, integrating over $\mathcal{C}_{ee} + \mathcal{C}_{xx}$ gives zero which guarantees detailed balance.
As before, we need to modify this expression to obtain a reasonable high temperature description of the spectrum, but we also wish to preserve the two properties just described.
To do this, we use the following approximation 
\begin{align}
 \mathcal{C}_{ee} &= \Gamma_{\nu\bar{\nu}} ( (4 \bar{u}_{ee} + \bar{u}_{xx}) (f_{\nu_e} - \rho_{ee}) + \bar{u}_{xx} f_{\nu_x} - \bar{u}_{ee} \rho_{ee} \nonumber\\
 &\quad + 4 {\rm Re}(\bar{u}_{ex}^* (f_{ex} - \rho_{ex})))\ ,\\
 \mathcal{C}_{xx} &= \Gamma_{\nu\bar{\nu}} ((4 \bar{u}_{xx} + \bar{u}_{ee}) (f_{\nu_x} - \rho_{xx}) + \bar{u}_{ee} f_{\nu_e} - \bar{u}_{xx}\rho_{xx} \nonumber\\
 &\quad+ 4 {\rm Re}( \bar{u}_{ex}^*(f_{ex} - \rho_{ex})) )\ ,\\
\mathcal{C}_{ex} &= \Gamma_{\nu\bar{\nu}}( 2 (\bar{u}_{ee} + \bar{u}_{xx} ) f_{ex}+ \bar{u}_{ex} (f_{\nu_e} + f_{\nu_x}))\nonumber\\
&\quad - \Gamma_{\nu\bar{\nu}} ( 3(\bar{u}_{ee} + \bar{u}_{xx} ) \rho_{ex} + 2 \bar{u}_{ex} (\rho_{ee} + \rho_{xx}))\ .
\end{align}
Here, $f_{\nu_\alpha}$ conserves number and energy of each flavor as before, and the replacement of $u_{\alpha\alpha} f_0$ by $f_{\nu_\alpha}$ guarantees that both lepton number and total number of neutrinos are conserved.

The amount of coherence is not conserved according to the expression for $\mathcal{C}_{ex}$. This is due to $\Tr{f_3\bar{f}_1}$ in the gain term. The physical interpretation is that neutrinos and antineutrinos in coherent states annihilate and the coherence is lost in the process.

Summing over the two contributions from neutrinos and antineutrino gives
\begin{align}
 \mathcal{C}_{ee} &= (\Gamma_{\nu\nu} (2u_{ee} + u_{xx}) + \Gamma_{\nu\bar{\nu}}( 4 \bar{u}_{ee} + \bar{u}_{xx}) ) (f_{\nu_e} - \rho_{ee}) \nonumber\\
&\quad + \Gamma_{\nu\bar{\nu}} (\bar{u}_{xx} f_{\nu_x} - \bar{u}_{ee}\rho_{ee}) + {\rm Re}((\Gamma_{\nu\nu} u_{ex}^* + 4 \Gamma_{\nu\bar{\nu}} \bar{u}_{ex}^*)(f_{ex} - \rho_{ex}))\ ,\\
\mathcal{C}_{xx} &= (\Gamma_{\nu\nu} (u_{ee} + 2 u_{xx}) + \Gamma_{\nu\bar{\nu}} (4 \bar{u}_{xx} + \bar{u}_{ee}) )(f_{\nu_x} -\rho_{xx})\nonumber\\ 
&\quad + \Gamma_{\nu\bar{\nu}} (\bar{u}_{ee} f_{\nu_e} - \bar{u}_{xx} \rho_{xx}) + {\rm Re}((\Gamma_{\nu\nu}u_{ex}^* + 4\Gamma_{\nu\bar{\nu}}\bar{u}_{ex}^*)(f_{ex} - \rho_{ex}))\ , \\
\mathcal{C}_{ex} &= \Gamma_{\nu\bar{\nu}} (\bar{u}_{ee} + \bar{u}_{xx}) (2f_{ex} - 3\rho_{ex})+ \Gamma_{\nu\bar{\nu}}\bar{u}_{ex} ((f_{\nu_e} + f_{\nu_x}) - 2(\rho_{ee} + \rho_{xx}))\nonumber\\
&\quad + \tfrac{3}{2} \Gamma_{\nu\nu} (u_{ee} + u_{xx}) ( f_{ex} - \rho_{ex}) + \tfrac{1}{2}\Gamma_{\nu\nu} u_{ex} ((f_{\nu_e}+f_{\nu_x}) - (\rho_{ee} + \rho_{xx}))\ .
\end{align}
In this form, the cancellations when integrating over momentum are seen very explicitly.
The corresponding expressions for antineutrinos are found by swapping $u\leftrightarrow\bar{u}$, $f_{\nu_\alpha} \leftrightarrow f_{\bar{\nu}_\alpha}$, $\rho\leftrightarrow \bar{\rho}$, except for $\Gamma_{\nu\nu}$ and $\Gamma_{\nu\bar{\nu}}$.
The coefficients for $\Gamma_{\nu\nu}$ and $\Gamma_{\nu\bar{\nu}}$ are $C_{\nu} = 0.122$ and $C_{\bar{\nu}} = 0.041$.

The effect of the third neutrino, $\nu_y$, which is assumed to do not oscillate can in principle also be taken into account. However, this induces loss terms in the equations, that are not balanced unless the number density of $\nu_y$ is also followed with a Boltzmann equation. The two terms in, e.g., $\mathcal{C}_{ee}$ that are most problematic are $2\Gamma_{\nu\bar{\nu}} (\bar{u}_{yy} f_{\nu_y} - \bar{u}_{ee}\rho_{ee})$ and $-\bar{u}_{ex}^* \rho_{ex}$. The first of these is direct annihilation to and from $\nu_y$, while the second term is annihilation of coherent states into $\nu_y$. In the case of full equilibrium and no conversions, there will be no effect, but if e.g.~$\nu_y$ has a different temperature or chemical potential, then there would be an untracked net flow of neutrinos in or out of the system. For now, we neglect this contribution and avoid the associated challenges.
Combining the terms from neutrinos, antineutrinos, electrons and positrons gives the collision term in \eref{eq:Cee} to (\ref{eq:Cem}), where we have replaced the term $d \; u_{ex} f_0$ by $ d \; f_{ex}$ for consistency.

\subsection{Validation of the collision term}

We have verified the accuracy of our approximation by comparing our results to the results of Ref.~\cite{Grohs:2015tfy} , where the full collision terms are calculated. To this purpose, we neglect flavor conversions in order to simplify the comparison.

\begin{figure}[h!]
 \centering
 \includegraphics[width=0.8\textwidth]{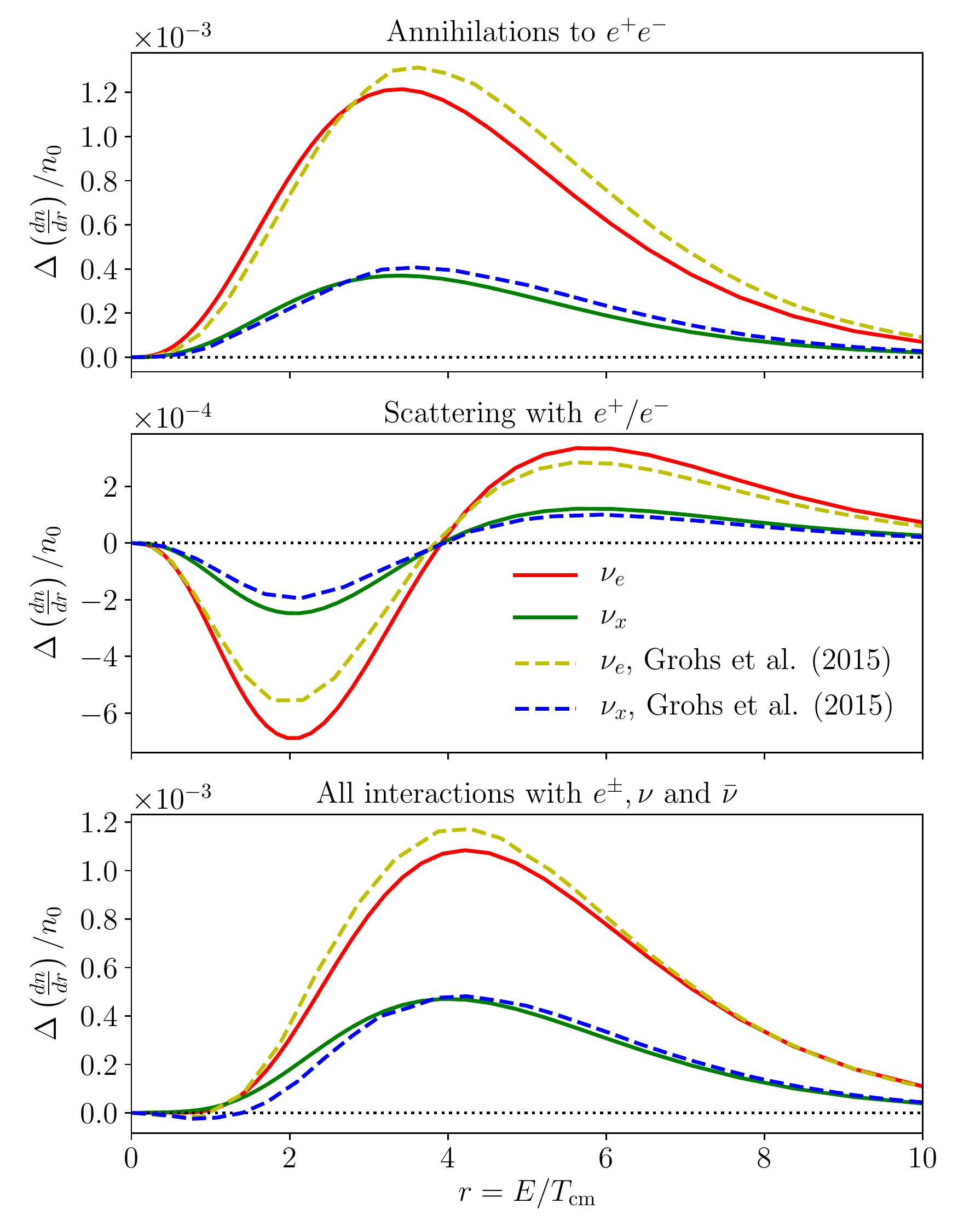}
 \caption{Absolute change in the neutrino/antineutrino number distribution as a function of $E/T_{\rm cm}$ at $T_\gamma \approx 0.1$~MeV. Each panel compares the results of Ref.~\cite{Grohs:2015tfy} to our results with the collision term approximated by Eqs.~(\ref{eq:Cee}) and (\ref{eq:Cmm}). {\it Top panel:} Spectrum from Fig.~7 of Ref.~\cite{Grohs:2015tfy} compared to our results computed using only the annihilation term. {\it Middle panel:} Spectrum from Fig.~8 of Ref.~\cite{Grohs:2015tfy} compared to our results obtained using only the scattering term. {\it Bottom panel:} Spectrum from Fig.~2 of Ref.~\cite{Grohs:2015tfy} and our results computed considering the scattering term and the annihilation term as well as the neutrino and antineutrino terms. Although we adopt an approximated collision term given to the challenges induced by our multi-angle setup, our results reproduce reasonably well the ones coming from the employment of the full collision term. }
 \label{fig:Grohs}
\end{figure}

We first consider annihilation into electrons and positrons by setting all front factors in Eqs.~(\ref{eq:Cee}) and (\ref{eq:Cmm}) to zero except for $\Gamma_{a,\alpha}$.
The functional form for the annihilation term is
\begin{equation}
 \label{eq:coll_anni}
 \Gamma_{a,\alpha} \left[ \left(\tfrac{T_\gamma}{T_{\rm cm}}\right)^4 f(T_\gamma,\mu_{\alpha}) - \rho_{\alpha\alpha} \right]\ .
\end{equation}
Compared to the usual relaxation approximation~\cite{Bell:1998ds, Dolgov:2001su, Hannestad:2015tea, Johns:2019hjl}, we have added $(T_\gamma/T_{\rm cm})^4$ to account for the fact that both electrons and positrons have a temperature higher than neutrinos.
The chemical potential $\mu_\alpha$ ensures the conservation of the lepton number [accounting for the factor of $(T_\gamma/T_{\rm cm})^4$], but allows the annihilation term to increase the number as well as the energy density of neutrinos.
In the upper panel of \fref{fig:Grohs}, we compare our results to those of Ref.~\cite{Grohs:2015tfy} for $\nu_e$ and $\nu_x$.
The agreement is reasonable considering our significantly simpler collision term. Using the approximation without the factor $(T_\gamma/T_{\rm cm})^4$ results in half the change.

Second, we take the scattering term. The functional form is
\begin{equation}
 \label{eq:coll_scat}
 \Gamma_{s,\alpha} \left(f(T_\gamma,\pi_{\alpha}) - \rho_{\alpha\alpha}\right),
\end{equation}
where the pseudo-chemical potential $\pi_{\alpha}$ ensures the conservation of the number density for each species and, hence, also the lepton number conservation. 
The results with all terms equal to zero except for $\Gamma_{s,\alpha}$ are shown in the second panel of \fref{fig:Grohs}.
The shapes of the spectral distortions that we find are again very close to the results obtained by using the full collision term, but the amplitude is larger by $\mathcal{O}(10\%)$. One source of discrepancy can be the neglected Pauli blocking, which can in itself give rise to deviations of that order of magnitude~\cite{Bell:1998ds,Hannestad:2015tea}. 

Finally, all collision terms are included in the results shown in the third panel of \fref{fig:Grohs}. The dominating effect comes from the heating through annihilations, but a suppression at small $E/T_{\rm cm}$ introduced by scatterings is also noticeable. The latter feature is also driven by the neutrino-neutrino interactions which tend to equilibrate the different neutrino species, while conserving the total neutrino number and energy density.
The neutrino term has the form: 
\begin{equation}
 \label{eq:coll_nunu}
(\Gamma_{\nu\nu} (2u_{\alpha\alpha} + u_{\beta\beta}) + \Gamma_{\nu\bar{\nu}}( 4 \bar{u}_{\alpha\alpha} + \bar{u}_{\beta\beta}) ) (f_{\nu_\alpha} - \rho_{\alpha\alpha}) + \Gamma_{\nu\bar{\nu}} (\bar{u}_{\beta\beta} f_{\nu_\beta} - \bar{u}_{\alpha\alpha}\rho_{\alpha\alpha})\ ;
\end{equation}
the first term proportional to $\Gamma_{\nu\nu}$ is due to neutrino-neutrino scattering while the terms with $\Gamma_{\nu\bar{\nu}}$ are due to neutrino-antineutrino interactions. The last term is specifically due to the annihilations $\nu_\beta \bar{\nu}_\beta \leftrightarrow \nu_\alpha \bar{\nu}_\alpha$, where both number and energy density can be transferred from one species to another one.

\subsection{Collision term for the homogeneous universe model with two angle bins}

In the homogeneous universe model with two angle bins, the collision terms need to differentiate between left and right moving neutrinos and antineutrinos. This is done by defining direction dependent equilibrium distributions. Furthermore, one might account for the angle dependence in the integrals by using different coefficients for parallel and antiparallel modes. However, for simplicity, we choose to use the same ``isotropic'' coefficients in both cases. With the assumption of ``isotropic'' coefficients, we find: 
\begin{eqnarray}
 \mathcal{C}_{ee,R} &=& \sum_{X\in\;\{R,L\}} \left[ \frac{\Gamma_{s,e}}{2} \left(f(T_\gamma,\pi_{e,X}) - \rho_{ee,R}\right) + \frac{\Gamma_{a,e}}{2} \left( \left(\tfrac{T_\gamma}{T_{\rm cm}}\right)^4 f(T_\gamma,\mu_{e,X}) - \rho_{ee,R} \right) \right.\nonumber\\
 && - \Gamma_G {\rm Re}\left( \bar{u}_{ex} \rho_{ex,R}^* \right) + \frac{1}{2}(\Gamma_{\nu\nu} (2u_{ee} + u_{xx}) + \Gamma_{\nu\bar{\nu}}( 4 \bar{u}_{ee} + \bar{u}_{xx}) ) (f_{\nu_e,X} - \rho_{ee,R}) \nonumber\\
 && \left.+ \frac{\Gamma_{\nu\bar{\nu}}}{2} (\bar{u}_{xx} f_{\nu_x,X} - \bar{u}_{ee}\rho_{ee,R}) + \frac{1}{2}{\rm Re}((\Gamma_{\nu\nu} u_{ex}^* + 4 \Gamma_{\nu\bar{\nu}} \bar{u}_{ex}^*)(f_{ex,X} - \rho_{ex,R}))\right]\ ,
 \end{eqnarray}
 \begin{eqnarray}
 \mathcal{C}_{xx,R} &=& \sum_{X\in\;\{R,L\}} \left[\frac{\Gamma_{s,x}}{2} (f(T_\gamma,\pi_{x,X}) - \rho_{xx,R}) + \frac{\Gamma_{a,x}}{2} \left( \left(\tfrac{T_\gamma}{T_{\rm cm}}\right)^4 f(T_\gamma,\mu_{x,X}) - \rho_{xx,R}\right) \right.\nonumber\\
 && - \Gamma_G {\rm Re}\left( \bar{u}_{ex} \rho_{ex,R}^* \right) + \frac{1}{2}(\Gamma_{\nu\nu} (u_{ee} + 2 u_{xx}) + \Gamma_{\nu\bar{\nu}} (4 \bar{u}_{xx} + \bar{u}_{ee}) )(f_{\nu_x,X} -\rho_{xx,R})\nonumber\\ 
 && \left.+ \frac{\Gamma_{\nu\bar{\nu}}}{2} (\bar{u}_{ee} f_{\nu_e,X} - \bar{u}_{xx} \rho_{xx,R}) + \frac{1}{2}{\rm Re}((\Gamma_{\nu\nu}u_{ex}^* + 4\Gamma_{\nu\bar{\nu}}\bar{u}_{ex}^*)(f_{ex,X} - \rho_{ex,R}))\right]\ ,
 \end{eqnarray}
 \begin{eqnarray}
 \mathcal{C}_{ex,R} &=& - D \rho_{ex,R} + d f_{ex} - C \bar{u}_{ex} (\rho_{ee,R} + \rho_{xx,R})\nonumber\\
 && + \sum_{X\in\;\{R,L\}} \left[ \frac{\Gamma_{\nu\bar{\nu}}}{2} (\bar{u}_{ee} + \bar{u}_{xx}) (2f_{ex,X} - 3\rho_{ex,R})
 + \frac{\Gamma_{\nu\bar{\nu}}}{2}\bar{u}_{ex} ((f_{\nu_e,X} + f_{\nu_x,X}) - 2(\rho_{ee,R} + \rho_{xx,R}))\right.\nonumber\\
 && + \tfrac{3}{4} \Gamma_{\nu\nu} (u_{ee} + u_{xx}) ( f_{ex,X} - \rho_{ex,R}) 
 + \tfrac{1}{4}\Gamma_{\nu\nu} u_{ex} ((f_{\nu_e,X}+f_{\nu_x,X}) - (\rho_{ee,R} + \rho_{xx,R}))\bigg]\ ,
\end{eqnarray}
where we have suppressed the dependencies on $\rho_R$, $\rho_L$, and $p$ for compactness.
The angular integrals are accounted for by multiplying the front factors by $1/2$ where appropriate. The chemical and pseudo chemical potentials $\mu_{\alpha,X}$ and $\pi_{\alpha,X}$ are determined from $\rho_{\alpha\alpha,X}$, and similarly for the neutrino equilibrium distributions $f_{\nu_\alpha,X}$ and $f_{\alpha\beta,X}$. The normalized energy densities are given by the average of the left and right moving mode: $u_{\alpha\beta} = \frac{1}{2}(u_{\alpha\beta,R}+u_{\alpha\beta,L})$. As for the Hamiltonian, the collision term for the left moving mode is obtained by $R\leftrightarrow L$.

\section{Quantum kinetic equations in the polarization vector formalism}
\label{app:QKE}
In this appendix, we derive the equation of motions in the formalism of the polarization vectors. Then, we show the transformation to the corotating frame, which speeds up the numerical solution of the equations of motion.
\subsection{Evolution equations}

Our implementation of the QKEs in LASAGNA allows for a non-zero lepton asymmetry. In order to follow small lepton asymmetries more accurately, we use the sums and differences of neutrino and antineutrino polarization vectors (see \eref{eq:poldef}):
\begin{equation}
 \label{eq:Ppm}
 \mathbf{P}^\pm_{R} = \mathbf{P}_{R} \pm \bar{\mathbf{P}}_{R}\ ,
\end{equation}
where the expressions for left moving modes are obtained by $R \leftrightarrow L$.
In addition, we find more convenient to replace $P_{0,R}^\pm$ and $P_{z,R}^\pm$ by
\begin{equation}
 \label{eq:Pab}
 P_{a,R}^\pm = P_{0,R}^\pm + P_{z,R}^\pm\ , \qquad P_{b,R}^\pm = P_{0,R}^\pm - P_{z,R}^\pm\ .
\end{equation}
In this work, $a=e$ stands for $\nu_e$ and $b=x$ (not to be confused with the $x$-component of the polarization vector) represents $\nu_x$. 
The polarization vectors are functions of the comoving momentum $r$, and time. With these variables, the term $-Hp \partial \rho_{R}/\partial p$ disappears, and the evolution equations for the polarization vectors are:
\begin{align}
 \label{eq:eompm}
 \frac{\partial}{\partial t} P_{a,R}^\pm &= V_{x,R}^s P_{y,R}^\pm + V_{x,R}^a P_{y,R}^\mp - V_{y,R}^s P_{x,R}^\pm - V_{y,R}^a P_{x,R}^\mp + \mathcal{C}_{aa,R} \pm \bar{\mathcal{C}}_{aa,R}\ ,\\
 \frac{\partial}{\partial t} P_{b,R}^\pm &= -V_{x,R}^s P_{y,R}^\pm - V_{x,R}^a P_{y,R}^\mp + V_{y,R}^s P_{x,R}^\pm + V_{y,R}^a P_{x,R}^\mp + \mathcal{C}_{bb,R} \pm \bar{\mathcal{C}}_{bb,R}\ ,\\
 \frac{\partial}{\partial t} P_{x,R}^\pm &= \frac{1}{2} V_{y,R}^s (P_{a,R}^\pm - P_{b,R}^\pm) + \frac{1}{2} V_{y,R}^a (P_{a,R}^\mp - P_{b,R}^\mp) - V_{z,R}^s P_{y,R}^\pm - V_{z,R}^a P_{y,R}^\mp \nonumber\\
 &\quad+ {\rm Re}(\mathcal{C}_{ab,R}) \pm {\rm Re}(\bar{\mathcal{C}}_{ab,R})\ ,\\
 \frac{\partial}{\partial t} P_{y,R}^\pm &= V_{z,R}^s P_{x,R}^\pm + V_{z,R}^a P_{x,R}^\mp - \frac{1}{2} V_{x,R}^s (P_{a,R}^\pm - P_{b,R}^\pm) - \frac{1}{2} V_{x,R}^a (P_{a,R}^\mp - P_{b,R}^\mp)\nonumber\\
 &\quad + {\rm Im}(\mathcal{C}_{ab,R})\pm {\rm Re}(\bar{\mathcal{C}}_{ab,R})\ ;
\end{align}
 the superscripts $s$ and $a$ denote the symmetric and antisymmetric parts of the potential with respect to neutrinos and antineutrinos:
\begin{align}
 &V_{x,R}^s(r) = -\frac{8\sqrt{2} G_{
\rm F} r T_{\rm cm}}{4 m_Z^2} \int \frac{dr'}{\pi^2} r' P_{x,L}^+(r') + \frac{\Delta m^2}{2 r T_{\rm cm}} \sin 2\theta\ ,\\
 &V_{x,R}^a(r) = \sqrt{2} G_{
\rm F} \int \frac{dr'}{2\pi^2} P_{x,L}^-(r') )\ ,
\end{align}
\begin{align}
 &V_{y,R}^s(r) = \sqrt{2} G_{
\rm F} \int \frac{dr'}{2\pi^2} P_{y,L}^+(r')\ ,\\
 &V_{y,R}^a(r) = -\frac{8\sqrt{2} G_{
\rm F} r T_{\rm cm}}{4 m_Z^2} \int \frac{dr'}{\pi^2} r' P_{y,L}^-(r')\ ,
\end{align}
\begin{align}
 &V_{z,R}^s(r) = -\frac{8\sqrt{2} G_{
\rm F} r T_{\rm cm}}{3m_W^2} u_e -\frac{\Delta m^2}{2 r T_{\rm cm}} \cos 2\theta\\ 
&\qquad\qquad-\frac{8\sqrt{2} G_{
\rm F} r T_{\rm cm}}{4m_Z^2}\int \frac{dr'}{\pi^2} r' \frac{1}{2}(P_{a,L}^+(r') - P_{b,L}^+(r')) \\
 &V_{z,R}^a(r) = \sqrt{2} G_{
\rm F} \int \frac{dr'}{2\pi^2} \frac{1}{2}(P_{a,L}^-(r') - P_{b,L}^-(r'))\ .
\end{align}
Notice that $V_{y,R}^s$ and $V_{y,R}^a$ depend on $P_{y,L}^-$ and $P_{y,L}^+$ respectively, which is opposite to the case for $V_{x,R}$ and $V_{z,R}$. This is due to a complex conjugation of the antineutrino density matrices, when the density matrices for neutrinos and antineutrinos are defined by similar equations.

The isotropic initial condition is set by using the equilibrium distributions for $P_a=P_0+P_z$ and $P_b=P_0-P_z$. The other components of the polarization vector are defined be means of the approximation~\cite{Hannestad:2012ky,Bell:1998ds}
\begin{equation}
 \label{eq:Pxyinit}
 P_x = \frac{V_x V_z}{\Gamma_D^2 + V_z^2}\ , \qquad P_y = -\frac{V_x \Gamma_D }{\Gamma_D^2+V_z^2}\ .
\end{equation}
Here $\Gamma_D$ is determined as $-{\rm Re}(\mathcal{C}_{ab,R})/P_{x,R}$, which means that $\Gamma_D \approx D$ from the collision term.

\subsection{Rotating frame}

At low temperatures, vacuum conversions dominate, and the solution is tracking the fast vacuum conversions of the lowest momentum states. However, the contribution from the vacuum term can be calculated analytically, and by going to a rotating frame, it can be eliminated from the equations.

The polarization vector in the flavor basis is $\mathbf{P} = (P_x,P_y,P_z)$. Since a rotation is a linear operation and neutrinos and antineutrinos experience the same vacuum term, $\mathbf{P}$ can also represent $\bar{\mathbf{P}}$ or $\mathbf{P}^\pm$. By adopting the following rotation matrix 
\begin{equation}
 \label{eq:Rmass}
 R_{\rm mass} =
 \begin{pmatrix}
 \cos(2 \theta) & 0 & -\sin(2 \theta) \\
 0 & 1 & 0 \\
 \sin(2 \theta) & 0 & \cos(2 \theta)
 \end{pmatrix}\ ,
\end{equation}
we go to the mass basis. The equations can then be written 
\begin{equation}
 \label{eq:Pmass}
 \frac{\partial}{\partial t} \mathbf{P} = (\mathbf{V}_{\rm mass} + \mathbf{V}_{\rm other}) \times \mathbf{P} + \mathcal{C}\ ,
\end{equation}
where $\mathbf{V}_{\rm mass}$ comes from the vacuum term, $\mathbf{V}_{\rm other}$ accounts for the remaining oscillation physics, and $\mathcal{C}$ is the collision term.
The same equations apply after rotating with $R_{\rm mass}$ since there is no time dependence. In the mass frame, $\mathbf{V}_{\rm mass} = V_{\rm mass} \mathbf{z}$, and $\mathbf{V}_{\rm mass}$ can be eliminated by rotating by
\begin{equation}
 \label{eq:Rvac}
 R_{\rm vac} =
 \begin{pmatrix}
 \cos(\phi) & -\sin(\phi) & 0 \\
 \sin(\phi) & \cos(\phi) & 0 \\
 0 & 0 & 1\\
 \end{pmatrix}\ ,
\end{equation}
where
\begin{equation}
\label{eq:Vvacphase}
 \phi = \int_{a_{\rm rot}}^a da' \frac{dt}{da'} V_{\rm mass}(a') = - \int_{a_{\rm rot}}^a da' \frac{\Delta m^2}{2 r T_{i} a_{i}} \;\frac{dt}{da'} \; a'\ ,
\end{equation}
$a_{\rm rot}$ is the scale factor were the solution in the rotating frame starts, and $T_i$ and $a_i$ are the initial temperature and scale factor.
Since we do not know the precise form of $dt/da'$ analytically, we solve the equation
\begin{equation}
 \label{eq:diffphi}
 \frac{d\phi_r}{da} = -\frac{\Delta m^2}{2 T_{i} a_{i}} \;\frac{a}{\dot{a}} = -\frac{\Delta m^2}{2 T_{i} a_{i}} \;\frac{1}{H}\ ,
\end{equation}
where $\phi_r = \phi r$, along with the QKE and the equation for $T_\gamma$.
In order to have a continuous transition from the non-rotating to the rotating frame, the polarization vector needs to be at the same point in the two frames when the transition occurs. This can be accomplished by rotating by $R_{\rm mass}$ after $R_{\rm vac}$ has been applied.

To summarize, we use the rotating basis $\mathbf{P}'$ which is related to $\mathbf{P}$ by
\begin{equation}
 \label{eq:Pprime}
 \mathbf{P}' = R_{\rm mass} R_{\rm vac} R_{\rm mass}^T \mathbf{P} = R \mathbf{P}\ .
\end{equation}
The full rotation matrix is
\begin{equation}
 \label{eq:Rmatrix}
 R =
 \begin{pmatrix}
 \cos\phi \cos^2 2\theta + \sin^2 2\theta & -\sin\phi \cos 2\theta & (\cos\phi-1) \cos 2\theta \sin 2\theta \\
 \sin\phi \cos 2\theta & \cos \phi & \sin\phi \sin 2\theta\\
 (\cos\phi - 1) \cos 2\theta \sin 2\theta & -\sin\phi \sin 2\theta & \cos\phi \sin^2 2\theta + \cos^2 2\theta 
 \end{pmatrix}\ .
\end{equation}
The equation for $\mathbf{P}'$ is
\begin{equation}
\label{eq:Primeevol}
 \frac{\partial}{\partial t} \mathbf{P}' = R \mathbf{V}_{\rm other} \times \mathbf{P} + R \mathcal{C}(\mathbf{P}) = R \mathbf{V}_{\rm other} \times (R^T \mathbf{P}') + R \mathcal{C}(R^T \mathbf{P}')\ .
\end{equation}
The speed gain is mainly present for low temperatures, but the overhead of going to the rotating frame is so small that LASAGNA becomes slightly faster even at $2$ and $4$~MeV.

\section{Growth of the neutrino-antineutrino asymmetry}
\label{app:nunubar}
In this appendix, we consider a very simple homogeneous and isotropic single-energy model (see e.g.~\cite{Hannestad:2006nj}) to explain the growth of the neutrino-antineutrino asymmetry through the linear stability analysis.

The linear stability analysis in \sref{sec:linear} considers the growth of the off-diagonal part of the density matrix. However, even if the off-diagonal part is different for neutrinos and antineutrinos, this cannot be measured and does not in itself imply that the densities are different.
In order to address this, we write the density matrices as\begin{equation}
  \rho = \frac{1}{2} \Tr{\rho} + \frac{1}{2}
  \begin{bmatrix}    s  - \delta & \epsilon \\ \epsilon^* & -s + \delta   \end{bmatrix} \;,\qquad
  \bar\rho = \frac{1}{2} \Tr{\bar\rho} + \frac{1}{2}
  \begin{bmatrix}    \bar{s} - \bar\delta & \bar{\epsilon}^* \\ \bar{\epsilon} & -\bar{s} + \bar\delta   \end{bmatrix} \;.
\end{equation}
Here $\epsilon$ and $\bar\epsilon$ are assumed to be small off-diagonal entries, while $\delta$ and $\bar\delta$ are small perturbations to the initial diagonal values $s$ and $\bar{s}$.
It is customary to ignore the mixing angle $\theta$ while performing linear stability analysis, but the growth of the neutrino-antineutrino asymmetry is sensitive to the vacuum mixing angle.
As for $\epsilon$ and $\bar\epsilon$, it is still a good approximation to neglect the vacuum mixing angle, and in the linear regime we find
\begin{equation}
  \label{eq:depsdt}
  \frac{d}{dt}  \begin{pmatrix} \epsilon \\ \bar\epsilon \end{pmatrix} = i
  \begin{pmatrix}
    \mu \bar{s} + \omega & -\mu s \\ \mu s & - \mu s - \omega
  \end{pmatrix}
  \begin{pmatrix} \epsilon \\ \bar\epsilon  \end{pmatrix} \;.
\end{equation}
We also notice that there is no feedback in the linear regime from the diagonals of the density matrix to the off-diagonals. Hence this growth is due to the well known  bipolar collective effects.

Proceeding to the equations for $\delta$ and $\bar\delta$, we assume that $s$ is constant and only $\delta$ and $\bar\delta$ are functions of time. Focusing on the asymmetry, we get
\begin{equation}
  \label{eq:deltaasym}
  \frac{d (\delta-\bar\delta)}{dt} = \omega \sin 2\theta \; [{\rm Im}(\epsilon) + {\rm Im}(\bar\epsilon)] \;.
\end{equation}
This equation clearly shows that a physical asymmetry between neutrinos and antineutrinos arises when the imaginary parts of the off-diagonal terms are not the negation of each other.
Hence, we need to determine the solutions for $\epsilon$ and $\bar\epsilon$.

In order to find linearly unstable solutions, we assume that
\begin{equation}
  \label{eq:eps_exp}
  \epsilon = Q e^{-i \Omega t}\ , \qquad \bar{\epsilon} = \bar{Q} e^{-i \Omega t}\ .
\end{equation}
With this ansatz, we can solve the equations when
\begin{equation}
  \label{eq:Omega}
  \Omega^\pm = \pm \sqrt{\omega ( 2 s \mu + \omega )}\;,
\end{equation}
which can be found as the eigenvalues of the coefficient matrix in \eref{eq:depsdt}.
The two eigenvalues correspond to the eigenvectors
\begin{equation}
  \label{eq:weq}
  \vec{w}^\pm = \begin{pmatrix} 1 + \frac{\omega - \Omega^\pm}{s \mu} \\ 1 \end{pmatrix} \;.
\end{equation}
Assuming IO ($\omega < 0$), we find that $\Omega^\pm$ are imaginary, and that the solution with $\Omega^+$ is growing exponentially. This growth is associated with $w^+$ meaning that $(Q,\bar{Q})^T$ has to be proportional to $w^+$.

We first consider an initial condition with symmetry between neutrinos and antineutrinos which we represent by the vector $\vec{v}_s = (1,1)^T$. In order to determine the component of $\vec{v}_s$ that will seed exponential growth, we project $\vec{v}_s$ on $\vec{w}^+$. 
For ${\rm Re}(\Omega^+) = 0$, this gives
\begin{equation}
  \label{eq:projsym}
  \begin{pmatrix}
    Q \\ \bar{Q}
  \end{pmatrix}_s = \frac{1}{2}
  \begin{pmatrix}
    1 + w^+_{\epsilon} w^{+*}_{\bar\epsilon}\\1 + w^{+*}_{\epsilon} w^+_{\bar\epsilon}
  \end{pmatrix}
  \;.
\end{equation}
Since $\Omega$ has no real part, the imaginary part of $\epsilon$ and $\bar\epsilon$ is determined by \eref{eq:projsym}. Here it is clear that the imaginary part has opposite signs for neutrinos and antineutrinos, and thus  the difference between $\delta$ and $\bar\delta$ will not grow.

We next consider an antisymmetric initial condition $\vec{v}_a = (1,-1)^T$. Projecting this on $\vec{w}^+$ and using ${\rm Re}(\Omega^+) = 0$, we get
\begin{equation}
  \label{eq:projanti}
  \begin{pmatrix}
    Q \\ \bar{Q}
  \end{pmatrix}_a = \frac{1}{2}
  \begin{pmatrix}
    1 - w^+_{\epsilon} w^{+*}_{\bar\epsilon}\\-1 + w^{+*}_{\epsilon} w^+_{\bar\epsilon}
  \end{pmatrix}
  \;.
\end{equation}
Here, the imaginary parts are equal for neutrinos and antineutrinos. Hence we expect to see a growing physical asymmetry. This is also confirmed by a numerical solution of the equations.

From this analysis it is clear that the exponential growth of $\delta-\bar\delta$ is not a new instability in itself, but it arises as part of the bipolar behavior of the system. In the context of the early universe, this effect has never been appreciated before since studies with approximate neutrino-antineutrino symmetry  have neglected the neutrino-neutrino term in the Hamiltonian. For core-collapse supernovae or compact binary merger remnants, such a highly symmetric initial condition is not realistic, and only the simplest studies have used it.

\bibliographystyle{JHEP}
\bibliography{literature}

\end{document}